\documentclass[10pt,sigconf]{acmart}

\usepackage{amsmath} 
\usepackage{balance}
\usepackage{booktabs} 
\usepackage{caption}
\usepackage{comment}
\usepackage{fancyvrb} 
\usepackage{graphicx} 
\usepackage{makecell} 
\usepackage{multirow} 
\usepackage{subfigure} 
\usepackage{soul}
\usepackage{xcolor}
\usepackage{xspace} 

\setcopyright{rightsretained}
\acmConference[SIGMOD'20]{ACM SIGMOD conference}{June 2020}{Portland, OR, USA}
\acmYear{2020}
\copyrightyear{2020}
\newcommand{\rewrite}[1]{{#1}}

\newcommand{\my}[1]{\textit{#1}}
\newcommand{\myindex}{\textit{SIRI}\xspace}
\newcommand{\mytree}{POS-Tree\xspace}
\newcommand{\myfullindex}{Structurally Invariant and Reusable Indexes\xspace}

\newcommand{\mpt}{MPT\xspace}
\newcommand{\nullNode}{\textit{null}\xspace}
\newcommand{\branchNode}{\textit{branch}\xspace}
\newcommand{\leafNode}{\textit{leaf}\xspace}
\newcommand{\extensionNode}{\textit{extension}\xspace}

\newcommand{\ceil}[1]{\left\lceil #1 \right\rceil}

\begin{document}

\title{Analysis of Indexing Structures for Immutable Data}




\author{
Cong Yue{\small $^{\dag}$},
Zhongle Xie{\small $^{\dag}$},
Meihui Zhang{\small $^{\S}$},
Gang Chen{\small $^{\ddag}$}, \\
Beng Chin Ooi{\small $^{\dag}$},
Sheng Wang{\small $^{\star}$},
Xiaokui Xiao{\small $^{\dag}$}
}
\affiliation{
$^{\dag}$\, National University of Singapore\\
$^{\S}$\, Beijing Institute of Technology\\
$^{\ddag}$\, Zhejiang University\\
$^{\star}$\, Alibaba Group\\
}


\begin{abstract}
In emerging applications such as blockchains and collaborative data analytics, there are strong demands for data immutability, multi-version accesses, and tamper-evident controls. 
To provide efficient support for lookup and merge operations,
three new index structures for immutable data, namely Merkle Patricia Trie (MPT), Merkle Bucket Tree (MBT), and Pattern-Oriented-Split Tree (\mytree), have been proposed.
Although these structures have been adopted in real applications, there is no systematic evaluation of their pros and cons in the literature, making it difficult for practitioners to choose the right index structure for their applications.

To alleviate the above problem,
we present a comprehensive analysis of the existing index structures for immutable data, and evaluate both their asymptotic and empirical performance.
Specifically, we show that MPT, MBT, and \mytree are all instances of a recently proposed framework, dubbed \my{Structurally Invariant and Reusable Indexes (SIRI)}.
We propose to evaluate the SIRI instances on their index performance and deduplication capability.
We establish the worst-case guarantees of each index,
and experimentally evaluate all indexes in a wide variety of settings.
Based on our theoretical and empirical analysis, we conclude that \mytree is a favorable choice for indexing immutable data.

\end{abstract}

\maketitle

\section{Introduction}
\label{sec:intro}

\noindent

Accurate history of data is required for auditing and tracking purposes in numerous practice settings. In addition, data in the cloud is often vulnerable to malicious tampering.
To support data lineage verification and  mitigate malicious data manipulation, data immutability is essential for applications, such as banking transactions and emerging decentralized applications (e.g., blockchain, digital banking, and collaborative analytics).
From the data management perspective, data immutability leads to two major challenges.

\begin{figure}[t]
\centering
\includegraphics[width=0.35\textwidth]{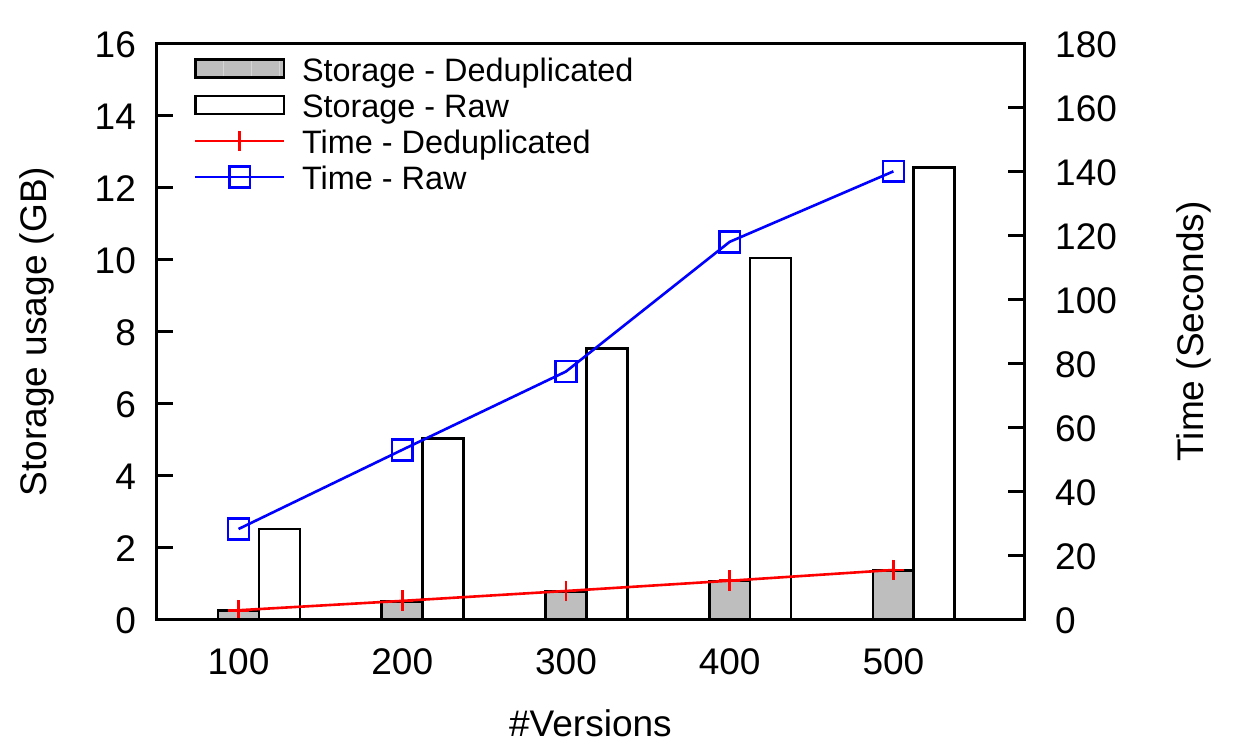}
\vspace{-2ex}
\caption{Data storage and transmission time improved by deduplication}
\label{fig:dedup_time_space}
\vspace{-2ex}
\end{figure}

First, it is challenging to cope with the ever-increasing volume of data caused by immutability.
An example is the sharing and storage of the data for healthcare analytics.
Data scientists and clinicians often make relevant copies 
of current and historical data in the process of data analysis, cleansing, and curation. 
Such replicated copies could consume an enormous amount of space and network resources.
To illustrate, let us consider a dataset that has 100,000 records initially, and it receives 1,000 record updates in each modification.
Figure~\ref{fig:dedup_time_space} shows the space and time required to handle the increasing number of versions\footnote{Run with  Intel(R) Xeon(R) E5-1620 v3 CPU and 1 Gigabit Ethernet card. }.
Observe that (i) the space and time overheads are significant if all versions are stored separately, and (ii) such overheads could be considerably reduced if we can {\it deduplicate} the records in different versions.

The second challenge is that in case a piece of data is tampered with (e.g., malicious manipulation of crypto-currency wallets or unauthorized modifications of patients' lab test data), we have to detect it promptly.
To address this challenge, the system needs to incorporate tamper-resistant techniques to support the authentication and recovery of data, to ensure data immutability.
Towards this end, a typical approach is to adopt cryptographic methods for tamper mitigation, which, however, considerably complicates the system design.


Most existing data management solutions tackle the above two challenges separately, using independent orthogonal methods.
In particular, they typically (i) ensure tamper evidence using cryptographic fingerprints and hash links~\cite{nakamoto:2009}, and (ii) achieve deduplication with delta encoding~\cite{maddox:2016,xu:2017}.
Such decoupled design incurs unnecessary overheads that could severely degrade the system performance.
For example, in state-of-the-art blockchain systems such as Ethereum~\cite{web:ethereum} and Hyperledger~\cite{web:hyperledger}, tamper evidence is externally defined and computed on top of the underlying key-value store (e.g., LevelDB~\cite{web:leveldb} or RocksDB~\cite{web:rocksdb}), which leads to considerable processing costs.
In addition, delta-encoding-based deduplication (e.g., in Decibel~\cite{maddox:2016}) requires a reconstruction phase before an object can be accessed, which renders data accessing rather inefficient. 

Motivated by the above issues, recent work~\cite{wang:2018,Wood:2014,web:hyperledger} has explored data management methods to provide native supports for both tamper evidence and deduplication features.
This results in three new index structures for immutable data, namely, \my{Merkle Patricia Trie (MPT)}~\cite{Wood:2014}, \my{Merkle Bucket Tree (MBT)}~\cite{web:hyperledger}, and \my{Pattern-Oriented-Split Tree (\mytree)}~\cite{wang:2018}.
To the best of our knowledge, there is no systematic comparison of these three index structures in the literature, and the characteristics of each structure are not fully understood.
This renders it difficult for practitioners to choose the right index structure for their applications. 

To fill the aforementioned gap,
this paper presents a comprehensive analysis of MPT, MBT, and \mytree.
Specifically, we make the following contributions: 
\begin{itemize}
\item We show that MPT, MBT, and \mytree are all instances of a recently proposed framework, named Structurally Invariant and Reusable Indexes (\myindex) \cite{wang:2018}.
Based on this, we identify the common characteristics of them in terms of tamper evidence and deduplication.

\item  We propose a benchmarking scheme to evaluate \myindex instances based on five essential metrics: their efficiency for four index operations (i.e., lookup, update, comparison, and merge), as well as their {\it deduplication ratios}, which is a new metric that we formulate to quantify each index's deduplication effectiveness.
We establish the worst-case guarantee of each index in terms of these five metrics. 

\item We experimentally evaluate all three indexes in a variety of settings.
We demonstrate that they perform much better than conventional indexes in terms of the effectiveness of deduplication.
Based on our experimental results, we conclude that \mytree is a favorable choice for indexing immutable data.
\end{itemize}

The rest of the paper is organized as follows.
Section~\ref{sec:background} presents the background and prior researches on the core application properties. Section~\ref{sec:siri} presents 
\myindex, along with an extended discussion on its significance and the explanation of three \myindex representatives.
A theoretical analysis is conducted in Section~\ref{sec:analysis} to reveal the operational bounds of \myindex while the experimental evaluation is reported in Section~\ref{sec:exp}.
We conclude this paper 
in Section~\ref{sec:conclu}.


\section{Related Work}
\label{sec:background}

\noindent

We first discuss the background and several primary motivations leading to the definition of \myindex.

\subsection{Versioning and Immutability}
\noindent

Data versioning has been widely employed for tolerating failures, errors, and intrusions,
as well as for analysis of data modification history.
ElephantFS~\cite{santry:1999} is one of the first-generation file systems with built-in multi-version support.
Successor systems like S4~\cite{strunk:2000}, CVFS~\cite{soules:2003}, RepareStore~\cite{zhu:2003} and OceanStore~\cite{kubiatowicz:2000} improve the early design by maintaining all versions in full scope and upon each update operation.
In databases, data versioning techniques are used for transactional data access.
Postgres~\cite{stonebraker:1987, bernstein:1983}, for example, achieved comparable performance to the database systems without versioning support.
Fastrek~\cite{chiueh:2005} enhanced Postgres with intrusion tolerance by maintaining an inter-transaction dependency graph based on the versioned data and relying on the graph to resolve data access conflicts.
Multi-versioning is also used to provide snapshot isolation in database systems \cite{Ports:2012, Berenson:1995} although such systems usually do not store the full history versions.
To directly access multi-versioned data,
a number of multi-version data structures can be applied from the literature,
such as multi-version B-tree~\cite{lanka:1991,rodeh:2008}, temporal hashing~\cite{Kollios:2002} and persistent data structures~\cite{driscoll:1989,okasaki:1999}.

Immutable data are becoming versatile in emerging applications.
For example, blockchains~\cite{web:ethereum, web:hyperledger, nakamoto:2009, anh:tkde, ruan:2019, blockbench} maintain immutable ledgers, which keep all historical versions of the system status.
Similarly, collaborative applications~\cite{bhardwaj:2015,web:googledocs} maintain the whole evolutionary history of datasets and derived analytic results, which enables provenance-related functionalities, such as tracking, branching, and rollback.
A direct consequence of data immutability is that all stored data are inherently multi-versioned upon being amended.
There exist a wide range of storage systems handling such data in either a linear manner, such as multi-version file systems~\cite{santry:1999, strunk:2000, soules:2003} and temporal databases~\cite{ahn:1986, salzberg:1999, tansel:1993}, or a non-linear manner, such as version control systems including git~\cite{web:git}, svn~\cite{web:svn} and mercurial~\cite{web:mercurial}, and collaborative management databases including Decibel~\cite{maddox:2016} and OrpheusDB~\cite{xu:2017}.
Git and Git-like systems are also used to manage the history and branches of datasets to achieve efficient query and space utilization~\cite{bhardwaj:2015,stonebraker:2012,web:gitfordata}.

\subsection{Data-Level Deduplication}
\label{subsec:dedup}
\noindent



Deduplication approaches have been proposed to reduce the overhead of storage consumption when maintaining multi-versioned data.
For example, Decibel~\cite{maddox:2016} uses {\em delta encoding},
whereby the system only stores the differences, called delta, between the new version and the previous version of data.
Consequently, it is effective to manage data versions when the deltas are small, despite the extra cost incurred during data retrieval for reconstructing the specified version of data.
However, it is ineffective in removing duplicates among non-consecutive versions or different branches of the data.
Though some algorithms choose a more precedent version that has the smallest differences as the parent to improve the efficiency of the deduplication, it involves additional complexity to reconstruct a version.


To enable the removal of duplicates among any data versions, {\em chunk-based deduplication} can be applied.
Unlike delta encoding, this approach works across independent objects.
It is widely used in file systems~\cite{paulo:2014, xia:2016}, and is a core principle of git.
In this approach, files are divided into {\em chunks}, each of which is given a unique identifier calculated from algorithms like collision-resistant hashing.
{\em chunks} with the same identifier can be eliminated.
Chunk-based deduplication is highly effective in removing duplicates for large files that are rarely modified.
In case an update leads to a change of all subsequent chunks, i.e., the boundary-shifting problem~\cite{eshghi:2005}, content-defined chunking~\cite{muthitacharoen:2001} can be leveraged to avoid expensive re-chunking.

\subsection{Tamper Evidence}
\label{subsec:tamper}
\noindent

Applications such as digital banking~\cite{web:webank} and blockchain~\cite{web:ethereum, web:hyperledger, nakamoto:2009} demand the system should maintain the accurate history of data, protect their data from malicious tampering, and trigger alerts when malicious tampering occur.
To serve such purposes, verifiable databases (i.e., Concerto~\cite{concerto:2017}, QLDB~\cite{web:qldb}) and blockchain services (i.e., Microsoft Azure Blockchain ~\cite{web:msbc}) often use cryptographic hash functions (e.g., SHA) and Merkle trees~\cite{merkle:1987} to verify the data integrity. \myindex, with the built-in support for tamper evidence, is a good candidate for the above systems.

A Merkle tree is a tree of hashes, where the leaf nodes are the cryptographic hashes calculated from blocks of data while the non-leaf nodes are the hashes of their immediate children.
The root hash is also called the ``digest'' of the data. 
To verify a record, it requires a ``proof'' of data, which contains the nodes on the path to the root.
The new root hash is recalculated recursively and equality is checked with the previously saved digest. 





\section{\myfullindex}
\label{sec:siri}
\noindent

\myfullindex (\myindex) are a new family of indexes recently proposed~\cite{wang:2018} to efficiently support tamper evidence and effective deduplication.

\subsection{Background and Notations}
\noindent
In addition to basic lookup and update operations, the ultimate goal of \myindex is to provide native data versioning, deduplication and tamper evidence features.
Consequently, data pages in \myindex must not only support efficient deduplication (to tackle the amount of replication arising from versioning) but also cryptographic hashing (to facilitate tamper evidence).


To better elaborate the \myindex candidates, we use the following notations in the remaining of this paper.
The indexing dataset is denoted as $\mathcal{D} = \{D_0, D_1, ..., D_n\}$ where $D_i$ represents its $i$-th version.
$\mathcal{I}$ is employed to represent \myindex structures, and $I$ stands for one of its instances. 
The key set stored in $I$ is set as $R(I) = \{r_1, r_2, ..., r_n\}$, where $r_i$ denotes the $i$-th key.
$P(I) = \{p_1, p_2, ..., p_n\}$ stands for the internal node set of $I$, where $p_i$ represents the $i$-th node.

\begin{figure}
	\centering
	\includegraphics[width=0.45\textwidth]{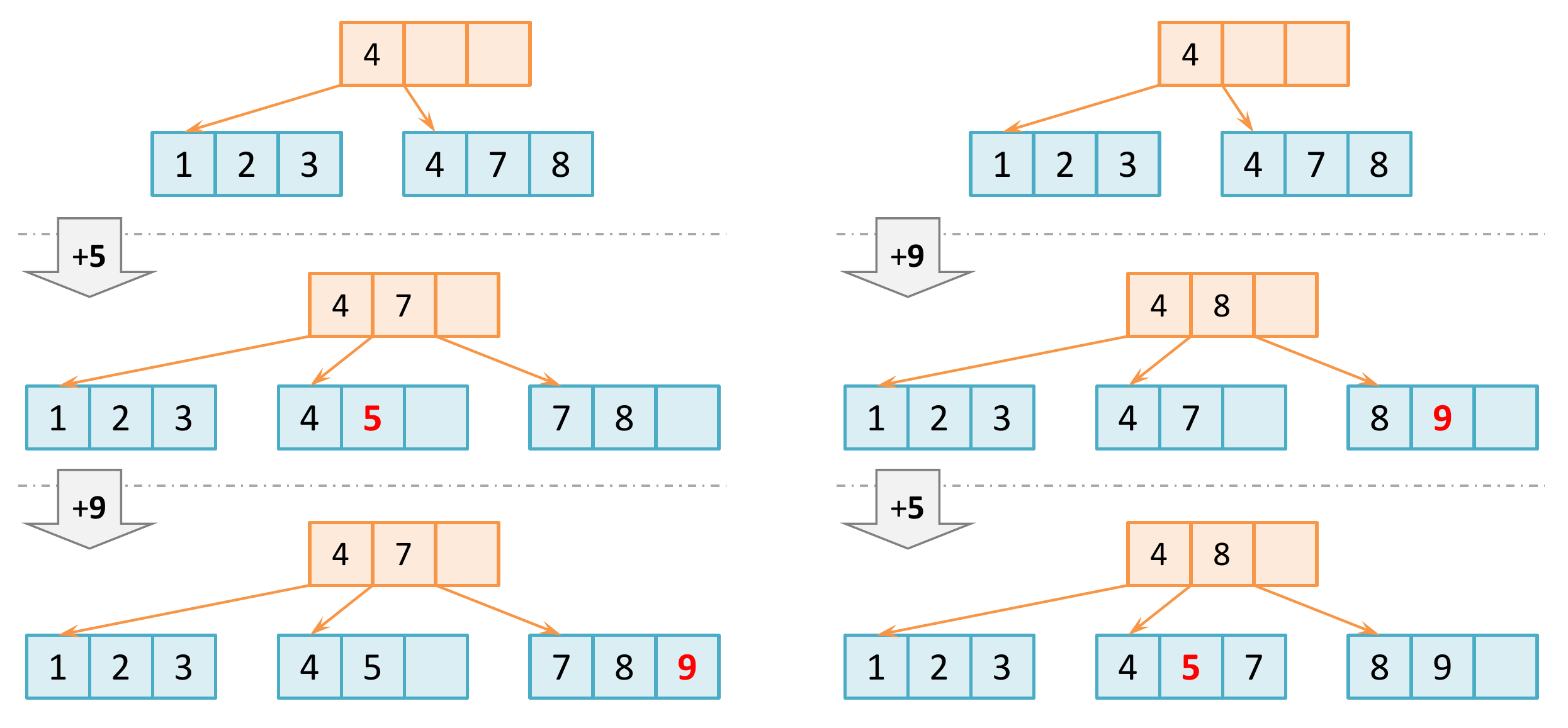}
	\vspace{-2ex}
	\caption{Two B$^+$-trees containing the same entries but with different internal structures~\cite{wang:2018}}
	\label{fig:btree}
	\vspace{-2ex}
\end{figure}

\subsection{Formal Definition}
\noindent

We provide a formal and precise definition of \myindex adapted from~\cite{wang:2018} as follows.

\begin{definition}
\label{def:siri}
An index class $\mathcal{I}$ belongs to \myindex if it has the following properties:
\begin{enumerate}
    \item \emph{Structurally Invariant}. If $I$ and $I'$ are two instances of $\mathcal{I}$, then
    $ P(I) = P(I') \iff R(I) = R(I').$
    \item \emph{Recursively Identical}. If $I$ and $I'$ are two instances of $\mathcal{I}$ and $R(I) = R(I') + r$, where $r \notin R(I')$, then
    $ |P(I) \cap P(I')| \gg |P(I) - P(I')|.$
    \item \emph{Universally Reusable}. For any instance $I$ of $\mathcal{I}$, there always exists node $p \in P(I)$ and another instance $I'$ such that $|P(I')| > |P(I)|)$ and $p \in P(I')$.
\end{enumerate}
\end{definition}

Definition~\ref{def:siri} states that \myindex must possess three properties.
The first property, \textit{Structurally Invariant}, ensures that the order of update operations does not affect the internal structure of the index, while the second property, \textit{Recursively Identical}, guarantees the efficiency when constructing a large instance from small ones. 
The third property, \textit{Universally Reusable}, secures that the nodes of the index could be shared among different instances.
In practice, these properties can be exploited to make \myindex time- and space-efficient.

\subsection{Extended Discussion}
\label{subsec:ext_disc}
\noindent

\textit{Recursively Identical} and \textit{Universally Reusable} are both aimed at making the pages share-able among various instances.
However, they focus on different aspects. 
The former attribute concentrates on providing performance improvement when designing the indexes -- updates do not bring in harmful impacts since the performance is often dominated by accessing a vast number of shared pages. 
The latter is to secure the theoretical boundary of \myindex's performance. 
The higher the ratio of shared pages each instance gets, the better performance \myindex could reach in terms of deduplication.
In the limiting case, where the dataset and indexing operations are infinite, every page in a \myindex instance could find its copy used by other instances.

It is non-trivial to construct a \myindex instance from conventional structures. 
Take the multi-way search tree as an example.
Such a structure is \textit{Recursively Identical} since only a small part of nodes is changed in the new version of the instance when an update operation is performed.
Further, the usage of copy-on-write implementation naturally enables node sharing among versions and branches.
Hence, it can be \textit{Globally Reusable} when applying this technique.
However, it may not be \textit{Structurally Invariant}.
Take B$^+$-tree as an example, Figure~\ref{fig:btree} illustrates that identical sets of items may lead to variant structures.
Meanwhile, hash tables are not \textit{Recursively Identical} when they require periodical reconstructions as the entire structure may be updated and none of the nodes can be reused.

Surprisingly, tries, or radix trees, can meet all the three properties with copy-on-write implementation.
Firstly, they are \textit{Structurally Invariant} since the position of the node only depends on the sequence of the stored key bytes and consequently, the same set of keys always leads to the same tree structure.
Secondly, being a multi-way search tree, they can be \textit{Recursively Identical} and \textit{Globally Reusable} as mentioned above.
However, they may end up in higher tree heights, leading to poor performance caused by increasing traversal cost, as shown in Section~\ref{sec:exp}.

Due to the appearance of the three properties, the adoption of the aforementioned data-level deduplication approaches can be seamlessly applied in index-level for \myindex.
The identical pages from different index instances for multiple versions of the data can be shared and therefore, the system can persist only one copy to save space.
Another benefit of applying index-level deduplication is that the system can access a version of the data directly from different indexes and data pages instead of experiencing a reconstruction phase from the deltas.
Overall, the nature of \myindex renders effective detection and removal of duplicates without prohibitive efforts, which conventional index structures can hardly offer.

\begin{figure}[t]
\centering
\includegraphics[width=0.47\textwidth]{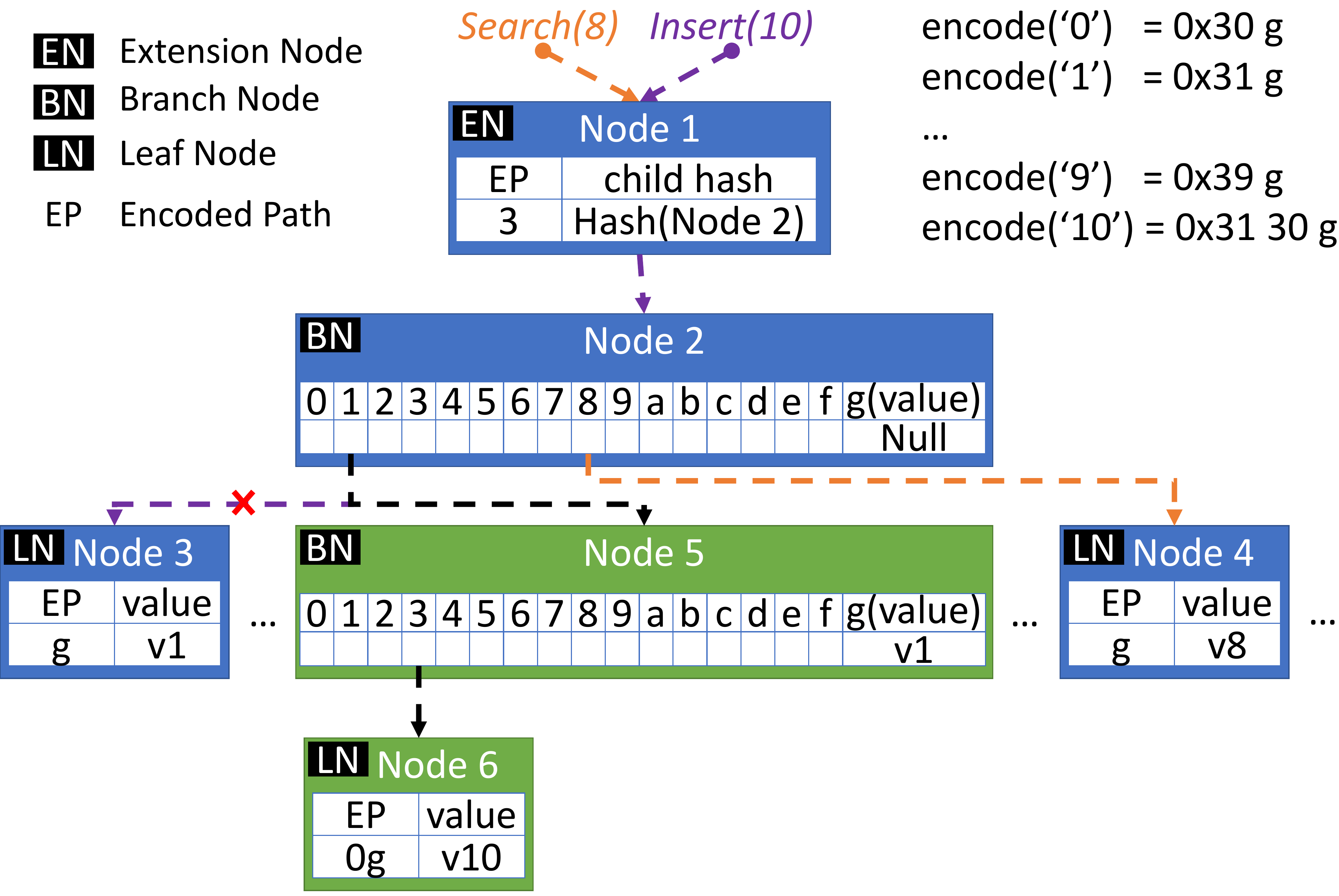}
\vspace{-2ex}
\caption{Merkle Patricia Trie (MPT)}
\label{fig:mpt}
\vspace{-3ex}
\end{figure}

\subsection{SIRI Representatives}
\noindent

In this section, we elaborate on the three representatives of \myindex, namely MPT, MBT, and \mytree.
As mentioned in section~\ref{subsec:ext_disc}, all representatives are \textit{Recursively Identical} and \textit{Globally Reusable} being multi-way search trees and leveraging a copy-on-write implementation of their nodes.
Meanwhile, they are \textit{Structurally Invariant} as stated in Section \ref{subsubsec:mpt}, \ref{subsubsec:mbt}, and \ref{subsubsec:pos}.

\subsubsection{Merkle Patricia Trie}
\label{subsubsec:mpt}
\noindent

Merkle Patricia Trie (MPT) is a radix tree with cryptographic authentication.
Similar to the traditional radix tree, the key is split into sequential characters, namely nibbles.
There are four types of nodes in \mpt, namely \branchNode, \leafNode, \extensionNode and \nullNode.
The structures of those nodes are illustrated in Figure~\ref{fig:mpt}:
(1) \branchNode node consists of a 16-element array and a value.
Each element, called ``branch'', of the array is indexing a corresponding child node and stores a nibble.
(2) \leafNode node contains a byte string, i.e., a compressed path called ``encodedPath'', and a value.
(3) \extensionNode node also contains encodedPath and a pointer to the next node.
(4) \nullNode node includes an empty string indicating that the node contains nothing.
Similar to Merkle Tree, the whole MPT can be rolled up to a single cryptographic hash for tamper evidence.
The most well-known usage of this data structure is in Ethereum~\cite{web:ethereum}, one of the largest blockchain systems in the world.

\textbf{Lookup.}
The lookup procedure for key ``8'' is illustrated in Figure~\ref{fig:mpt}.
The key is first encoded as ``0x38 g''.
Then, each character of the encoded key is used to match with the encodedPath in an extension node, or to select the path in a branch node, from left to right.
For this example, the first character ``3'' matches the root node's encodedPath, therefore, it navigates to its child, Node 2.
Then it takes the branch ``8'' since ``8'' equals to the second character in the encoded key.
Finally, the traversal reaches the leaf node and ends with the value ``v8'' output. 

\textbf{Insert.}
To insert data in MPT, the index first locates the position of the given key as in the lookup operation.
Once it reaches a \nullNode node, a leaf node containing the remaining part of the encoded key and the value is created.
For example, in Figure~\ref{fig:mpt}, when we insert key ``1'' (``0x31 g''), if branch ``1'' in Node 2 is empty, a new node (``g'', v1) is created and pointed by branch ``1''.
In case there is a partial match at extension node, a new branch node at diverging byte is created, appended with original and new child.
The insertion of key ``10'' in the figure can illustrate this procedure, where the path is diverged at Node 3.
Hence, Node 3 is replaced by Node 5 with a newly created Node 6 attached.

\subsubsection{Merkle Bucket Tree}
\label{subsubsec:mbt}
\noindent

\begin{figure}[t]
\centering
\includegraphics[width=0.48\textwidth]{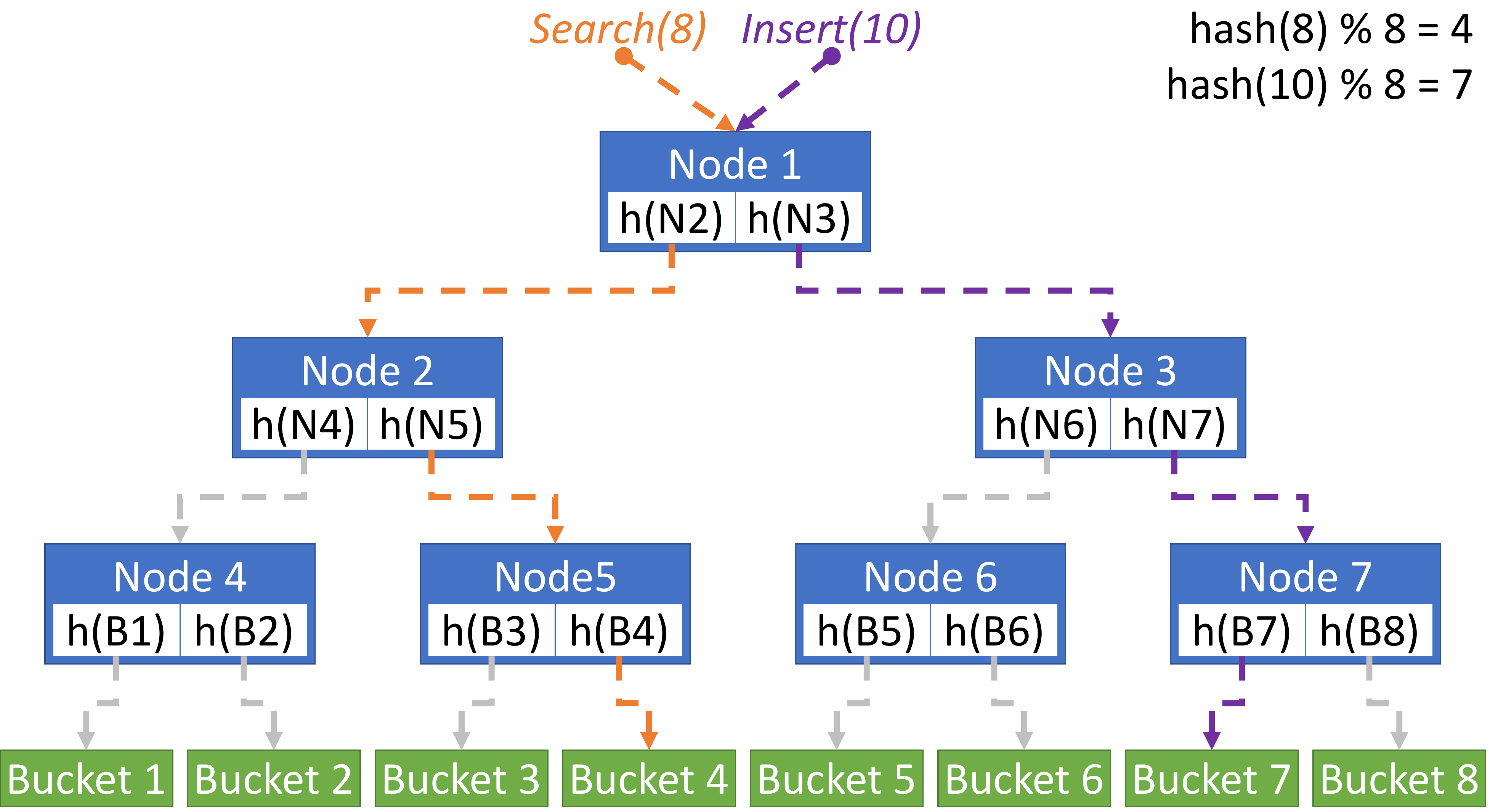}
\caption{Merkle Bucket Tree (MBT)}
\label{fig:mbt}
\vspace{-3ex}
\end{figure}

Merkle Bucket Tree (MBT) is a Merkle tree built on top of a hash table as shown in Figure~\ref{fig:mbt}.
The bottom most level of MBT is a set of buckets and the cardinality of the bucket set is called \textit{capacity}.
Data entries are hashed to these buckets, and the entries within each bucket are arranged in sorted order.
The internal nodes are formed by the cryptographic hashes computed from their intermediate children.
The number of children an internal node has is called \textit{fanout}.
In MBT, \textit{capacity} and \textit{fanout} are pre-defined and cannot be changed in its life cycle.

\textbf{Lookup.}
To perform an MBT index lookup, we first calculate the hash of the target key and obtain the index of the bucket where the data resides.
Due to the copy-on-write restrictions, we are unable to fetch the bucket directly and hence we then use the bucket number to calculate the traversal path from the root node to the leaf node.
The calculation is generally a trivial reverse simulation of the complete multi-way search tree search algorithm.
For example in Figure~\ref{fig:mbt}, key ``8'' falls into Bucket 4 after the hashing,
and we accordingly get all node index on the path starting from the leaf.
Finally, we follow the path to reach the bucket.
The records in the bucket are scanned using binary search to find the target key after the retrieval of the bucket node.

\textbf{Insert.}
The insert operation of MBT undergoes similar procedures.
It first performs a lookup to check the existence of the target key.
For example, the inserting key ``10'' falls into Bucket 7.
Then Bucket 7 is fetched following the lookup process, and the key is inserted to Bucket 7 in ascending order. 
Finally, the hashes of the bucket and the nodes are recalculated recursively.

The design of MBT undoubtedly takes the advantages of Merkle tree and the hash table.
On the one hand, MBT offers tamper evidence with a low update cost since only the set of nodes lying on the lookup path needs to be recalculated. 
On the other hand, the data entries can be evenly distributed due to the nature of the hash buckets in the bottom level.


\subsubsection{Pattern-Oriented-Split Tree}
\label{subsubsec:pos}
\noindent

\begin{figure}
\centering
\includegraphics[width=0.45\textwidth]{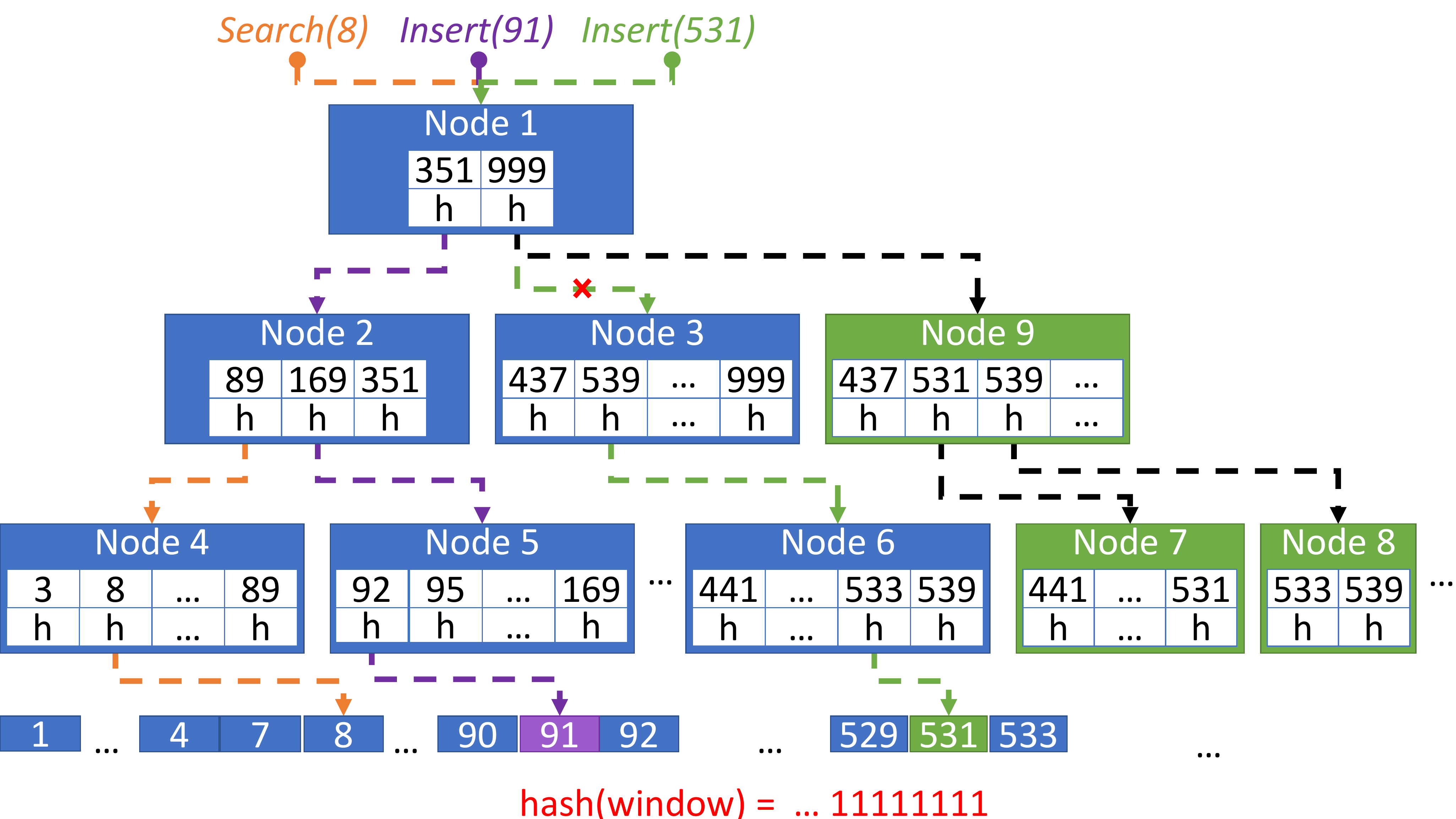}
\vspace{-2ex}
\caption{Pattern-Oriented-Splitting Tree (\mytree)}
\label{fig:pos-tree}
\vspace{-3ex}
\end{figure}

Pattern-Oriented-Split Tree (\mytree) is a probabilistically balanced search tree proposed in~\cite{wang:2018}.
The structure can be treated as a customized Merkle tree built upon pattern-aware partitions of the dataset, as shown in Figure~\ref{fig:pos-tree}.
The bottom most data layer is an ordered sequence of data records.
The records are partitioned into blocks using a sliding-window approach and such blocks form the leaf nodes.
That is, for a byte sequence within a fixed-sized window, starting from the first byte of the data, a Rabin fingerprint is computed to match a certain boundary pattern.
An example pattern can be the last 8 bits of Rabin fingerprint equaling to ``1''.
The window shifts forward to repeat the process until it finds a match, where the node boundary is set to create the leaf node.
The internal layers are formed by a sequence of split keys and cryptographic hashes of the nodes in the lower layer.
Since the contents in the internal layers already contain hash values, we directly use the hashes to match the boundary pattern instead of repeatedly computing the hashes within a sliding window.
Such strategy improves the performance of \mytree by reducing the number of hash computations, while preserving the randomness of chunking. 

\textbf{Lookup.}
The lookup procedure of \mytree is similar to B$^+$-tree.
Starting from the root node, it performs binary search to locate the child node containing the target key.
When it reaches the leaf node, a binary search is performed to find the exact key.
As the example shown in Figure~\ref{fig:pos-tree}, the key ``8'' is fetched through the orange path. It goes through Node 2, which has a key range of (-$\infty$, 351], and Node 4, which has a key range of (-$\infty$, 89].

\textbf{Insert.}
To perform an insert operation, \mytree first finds the position of the inserting key and then inserts it into the corresponding leaf node. 
Next, it starts the boundary detection from the first byte of the leaf node, and stops when detecting an existing boundary or reaching the last byte of the layer.
For example, when insert key ``91'' into the tree shown in Figure~\ref{fig:pos-tree}, a boundary detection is performed from Node 5. 
It ends upon reaching the existing boundary of Node 5.
Another instance in the figure is the insertion of key ``531''. 
A new boundary is found at element 531, and the traverse stops when finding the existing boundary of Node 6.
Therefore, Node 6 splits into Node 7 and Node 8, and the new split keys are propagated to the parent node.

The pattern-aware partitioning of \mytree enhances the deduplication capabilities, and making the structure of the tree depending only on the data held.
Such \textit{Structurally Invariant} property supports efficient \textit{diff} and \textit{merge}. Moreover, the B$^+$-tree-like node structure enables efficient indexing by comparing the split keys to navigate the paths.

\section{Theoretical Analysis}
\label{sec:analysis}
\noindent

\begin{table}[!t]
    \centering
    \caption{Notation table}
    \label{table:notation}
    \begin{tabular}{|l|l|}
        \hline
        Symbol & Description \\
        \hline
        $N$ & The total number of records\\
        \hline
        $m$ & The fanout of \mytree and MBT\\
        \hline
        $B$ & The capacity of MBT (Number of buckets)\\
        \hline
        $L$ & The key length of a record\\
        \hline
        $\delta$ & The number of different records \\
        & between two versions\\
        \hline
        $\alpha$ & The ratio of the number of different \\
        &records and the number of total records\\
        \hline
        $r$ & The average size of a record\\
        \hline
        $c$ & The size of cryptographic hash value\\
        \hline
    \end{tabular}
    \vspace{5pt}
\end{table}

In this section, we provide a comprehensive theoretical analysis of the three \myindex representatives discussed previously.
We first calculate the theoretical bounds for common index operations like lookup and update,
as well as complex operations needed by emerging applications including diff and merge.
In addition,
we define the \my{deduplication ratio} as a metric
for measuring the efficiency of deduplication provided by {\myindex}s.



In the following subsections, $N$ denotes the maximum capacity of distinct keys in an index; $L$ denotes the maximum length of a key string; $B$ denotes the number of buckets in MBT. $m$ denotes the expected number of entries per page/node in MBT or \mytree. $\delta$ denotes the different records between two instances.
$r$ denotes the average storage size for a sole record.

\subsection{Operation Bounds}
\label{subsec:operation_bounds}

In this section, the bounds of common operations are calculated accordingly.


\subsubsection{Index Lookup}
\label{subsubsec:lookup}
\noindent

We first evaluate the lookup complexity of the three candidates.






\begin{itemize}

\item \textbf{MPT} -- The traversal upon MPT is the same as a normal radix tree with path compaction. The computing bound for lookup is the maximum between $O(L)$ and $O(\log_m N)$.
Since $L$ is often larger than $\log_m N$ in the real systems ($L$ equals to 64-byte in Ethereum's zeroith variable), the lookup complexity in MPT is $O(L)$ in most of the time.

\item \textbf{MBT} -- Unlike other structures having constant leaf node scanning time, the size of MBT's leaf node is $\frac{N}{B}$ and it therefore costs $O(\log_2 \frac{N}{B})$ with binary search to scan the node.
As a result, the total lookup cost, consisting of node traversing part and leaf node scanning part, is $O(\log_m B + \log_2 \frac{N}{B})$.

\item \textbf{\mytree}\ -- Being a probabilistically balanced search tree, the complexity of \mytree is $O(\log_{m} N)$.

\end{itemize}

\subsubsection{Index Update}
\noindent



In all candidates, an update operation firstly incurs a lookup for the updating keys, and then leads to the creation of the copy of affected nodes and the calculation of their hash values.
In our analysis, we treat the number of entries in each node $m$, the length of each record $r$ and the length of hashed value $len(h)$ as constant values.
That is to say, the size of internal nodes $m \cdot len(h)$, and the size of leaf nodes $m \cdot r$ are constant unless explicitly stated.
Therefore, the cost of new node creation and the crypto-hash function is constant, and the cost of the update mainly depends on the cost of the lookup.

The calculation and comparison among the three candidates are listed below:




\begin{itemize}
    \item \textbf{MPT} -- $\max(O(L), O(\log_m N))$.
    In most cases, the complexity of the update in MPT is $O(L)$.
    
    \item \textbf{MBT} -- For leaf nodes, as the size increases linearly with $N$, the complexity of hash function and node copying is $O(\frac{N}{B})$.
    Hence, the complexity of the update in MBT is $O(\log_m B + \frac{N}{B})$.
    
    \item \textbf{\mytree}\ -- Similar to crypto-hash function, the cost of the rolling hash function p for detecting node boundary is also constant. This results in the update complexity of $O(\log_{m} N)$.
\end{itemize}


\subsubsection{Indexes Diff}
\noindent


\my{Diff} is the operation that compares two index instances.
It returns all records that are either present in only one index or different in both indexes. 
Therefore, \my{Diff} can be seen as multiple lookups in a naive implementation of the three candidates.
The following bounds are calculated under this assumption.
We directly give the results due to its triviality.

\begin{itemize}

\item \textbf{MPT} -- $O(\delta \cdot L)$ or $O(\delta \cdot \log_m N)$. As discussed previously, in most cases the complexity is the former.

\item \textbf{MBT} -- $O(\delta \cdot (\log_{m} B + \frac{N}{B}))$ 

\item \textbf{\mytree}\ -- $O(\delta \cdot \log_{m} N)$ 

\end{itemize}

\subsubsection{Indexes Merge}
\noindent


\my{Merge} is the operation that combines all records from either indexes.
The entire process of \my{Merge} contains two steps.
The first step is to do a \my{Diff} operation between the instance to merge and the original instance, mark the different pages/nodes.
The second step is to merge all the different nodes into the original instance.
If there exist conflicts, namely a key in both instances with different values, the process must be interrupted and a selection strategy must be given by the end user to continue.
The following calculation is based on the worst case when the merge process can be finished without interruption.
Since the second step of the merge process is treated as $O(1)$ operations in our analysis, the complexity of the merge is dominated by the ``diff'' operation in the first step.
\begin{itemize}

\item \textbf{MPT} -- $O(\delta \cdot L )$ or $O(\delta \cdot \log_m N)$. In most cases, the complexity should be $O(\delta \cdot L )$.

\item \textbf{MBT} -- $O(\delta \cdot (\log_{m} B + \frac{N}{B}))$ 

\item \textbf{\mytree}\ -- $O(\delta \cdot \log_{m} N)$ 

\end{itemize}

In the worst case, MPT has higher tree height than a balanced search tree, i.e., L > $O(\log_m N)$, and therefore performs worse than \mytree.
For MBT, the traverse cost $\log_m B$ is lower than other structures when in assumption of $B < N$ while the node scanning time $\log_2 \frac{N}{B}$ and creation time $\frac{N}{B}$ are dominating when $N >> B$.
We can conclude from the table that \mytree is efficient in general cases, while MBT is a good choice when the dataset maintains a proper $\frac{N}{B}$ ratio.

\subsection{Deduplication Ratio}
\noindent

Persistent (or immutable) data structures demand a large amount of space
for maintaining all historical versions of data.
To alleviate space consumption pressure,
the feasibility of detecting and removing duplicated data portions plays a critical role.
In this section, we aim to quantify the effectiveness of such properties in indexes by defining a measurement called \my{deduplication ratio}. 


\subsubsection{Definition}
\noindent

Suppose there is a set of index instances $S = \{I_1, I_2, ... I_k\}$,
and each $I_x$ is composed of a set of pages $P_x$.
The byte size of a page $p$ is denoted as $byte(p)$,
we can derive byte count of set $P$ as:
$$byte(P) = \sum_{p \in P} byte(p).$$
The \my{deduplication ratio} $\eta$ of $S$ is defined as follows:
$$ \eta (S) = 1 - \frac{byte(P_1 \cup P_2 \cup ... \cup P_k)}{byte(P_1) + byte(P_2) + ... + byte(P_k)},$$
or
$$ \eta (S) = 1 - \frac{byte( \bigcup_{i = 1}^{k} P_i )}{\sum_{j = 1}^{k} byte(P_j) }.$$


The $\eta$ quantifies the effectiveness of page-level data deduplication (i.e., sharing) among related indexes.
It is the ratio between the overall bytes that can be shared between different page sets and the total bytes used for all the page sets.
With a 
high $\eta$,
the storage is capable of managing massive ``immutable''
data versions without bearing space consumption pressure.
In the following subsections, we will use this metric to evaluate the three candidates accordingly.

\subsubsection{Continuous Differential Analysis}
\label{subsec:conti_analy}
\noindent





In this part, we analyze a simple case that $S$ consists $n$ sequentially evolved indexes,
i.e., the $i_{th}$ instance is derived from the $i-1_{th}$ instance.
Each instance $S_i$ can be represented as a page set $P_i$ or a record set $R_i$.
The analysis of more complicated scenarios is treated as our future work.
To ease our analysis, we assume that each instance differs its predecessor by ratio $\alpha$ of a continuous key range $\delta$, such that:
$$|D_i| = \alpha \cdot |R_{i-1}|,$$
$$\forall k \in (R_i - D_i) \cup (R_{i-1} - D_i), (k < min(D_i) \vee k > max(D_i))$$
where $|X|$ denotes the record count in set $X$,
and $min/max$ denotes the minimum/maximum key in a set.

In the following analysis, we consider two scenarios:

\begin{itemize}

\item Insertion of new records.
$$|R_i| = (1 + \alpha) \cdot |R_{i-1}|.$$

\item Update of existing records.
$$|R_i| = |R_{i-1}|.$$

\end{itemize}

\textbf{Merkle Bucket Tree.}
Since in MBT, the bucket size depends on the number of contained records, i.e.,
$$E = \frac{N}{B}.$$
We denote the number of affected nodes on level $x$ in MBT as $N_x$.
Hence, the number of buckets (the leaf level) affected by $\alpha$ differential is expressed as:
$$N_0 = \alpha \cdot B.$$



We can roughly summarize the total number of affected tree nodes, $N_{tree}$, as following: \begin{equation*}
\begin{split}
N_{tree} & = \ceil{\frac{B}{m}} + \ceil{\frac{B}{m^2}} + ... + \ceil{\frac{B}{m^{\log_{m}B}}} \\
& \approx \frac{B}{m} + \frac{B}{m^2} + ... + \frac{B}{m^{\log_{m}B}} \\
& = \frac{B-1}{m-1}.
\end{split}
\end{equation*}

The number of affected nodes in the continuous update can be calculated as:
\begin{equation*}
\begin{split}
{\sum_{i = 1}^{\log_{m}B} N_i } & = \ceil{\alpha \cdot \frac{B}{m}} + \ceil{\alpha \cdot \frac{B}{m^2}} + ... + \ceil{\alpha \cdot \frac{B}{m^{\log_{m}B}}} \\
& \approx \alpha \cdot \frac{B}{m} + \alpha \cdot \frac{B}{m^2} + ... + \alpha \cdot \frac{B}{m^{\log_{m}B}} \\
& = \alpha \cdot \frac{B-1}{m-1}.
\end{split}
\end{equation*}

Thus, the deduplication ratio for MBT is:
\begin{equation*}
\begin{split}
    \eta(MBT) & = 1 - \frac{\textrm{bytes in modified nodes}}{\textrm{total bytes}} \\
    & = 1 - \\ &\frac{\alpha \cdot N \cdot r + m \cdot c \cdot {\sum_{i = 1}^{\log_{m}B} N_i } + N \cdot r + m \cdot c \cdot N_{tree}}{2 \cdot (N \cdot r + m \cdot c \cdot N_{tree})} \\
    & \approx \frac{1}{2} - \frac{\alpha \cdot N \cdot r + m \cdot c \cdot \alpha \cdot \frac{B-1}{m-1} }{2 \cdot (N \cdot r + m \cdot c \cdot \frac{B-1}{m-1})} \\
    & = \frac{1}{2} - \frac{\alpha}{2}.
\end{split}
\end{equation*}

Surprisingly, the deduplication ratio is highly related to $\alpha$ and has no direct connection with $B$ according to the analysis result.

\textbf{Merkle Patricia Trie.}
In a simple MPT without any path compaction optimization, we have:


\begin{equation*}
\begin{split}
    \eta(MPT) &= 1 - \frac{\textrm{bytes in modified nodes}}{\textrm{total bytes}}\\
    &= \frac{1}{2} - \frac{\alpha \cdot N \cdot (L \cdot c + r)}{2 \cdot (N \cdot r + N \cdot \bar{L} \cdot c)},
\end{split}
\end{equation*}
which indicates that

$$\eta(MPT) \geq \frac{1}{2} - \frac{\alpha}{2}\ (L \geq \bar{L})$$,
$$\eta(MPT) \leq \frac{1}{2} - \frac{\alpha}{2}\ (L \leq \bar{L})$$

Inferred from the result, $\eta(MPT)$ is affected by the distribution of stored keys since the length of the average length of the keys highly relate to the final deduplication ratio.
In detail, the relationship between $L$ and $\bar{L}$ determines whether the deduplication ratio of MPT is greater than or less than that of MBT.

\textbf{\mytree.}
Similar to MBT, the calculation is as follows:

\begin{equation*}
\begin{split}
\eta(POS) & = 1 - \frac{\textrm{bytes in modified nodes}}{\textrm{total bytes}} \\
 & \approx \frac{1}{2} - \frac{\alpha \cdot N \cdot r + m \cdot c \cdot {\sum_{j = 1}^{\log_{m}\frac{N}{f}}N_j } }{2 \cdot (N \cdot r + \frac{\frac{N}{f}-1}{m - 1} \cdot m \cdot c)} \\
 & \approx \frac{1}{2} - \frac{\alpha \cdot N \cdot r + m \cdot c \cdot \alpha \cdot \frac{\frac{N}{f}-1}{m-1}}{2 \cdot (N \cdot r + \frac{\frac{N}{f}-1}{m - 1} \cdot m \cdot c)} \\
 & = \frac{1}{2} - \frac{\alpha}{2}.
\end{split}
\end{equation*}

If we compare the analysis results of the three representatives, we can conclude that MPT has the best deduplication ratio under proper query workloads and datasets (meaning $L \geq \bar{L}$).
Meanwhile, \mytree and MBT have equal bound for the deduplication ratio in this setting.

\section{Experimental Benchmarking}
\label{sec:exp}
\noindent 

\begin{table}[!t]
    \centering
    \caption{Parameter table for experiments}
    \label{table:para}
    \begin{tabular}{|l|l|}
        \hline
        Parameter & Value \\
        \hline
        Dataset size($10^4$) & 1, 2, 4, 8, \underline{16}, 32, 64, \\
        & 128, 256 \\
        \hline
        Batch size($10^3$) & 1, \underline{2}, 4, 8, 16\\
        \hline
        Overlap Ratio & \underline{0}, 10, 20, 30, 40, 50, 60,\\
        & 70, 80, 90, 100\\
        \hline
        Write Ratio(\%) & \underline{0}, 50, 100\\
        \hline
        Zipfian parameter {$\theta$} & \underline{0}, 0.5, 0.9 \\
        \hline
    \end{tabular}
    \vspace{5pt}
\end{table}

In this section, we evaluate three \myindex representatives, namely \mytree, MBT and MPT, through different experiments.
First, the throughput and the latency of the indexes are measured to have an overview of how these structures perform in general cases.
Second, the storage consumption, the deduplication ratio and the node sharing ratio are evaluated to investigate the space efficiency among the candidates.
Third, a breakdown analysis is given to show how each \myindex property affects the performance of the index.
Finally, we integrate the structures in an existing database management system, Forkbase~\cite{wang:2018}, to show how \myindex structures behave in real applications.

Our experiments are conducted on a server with Ubuntu 14.04, which is equipped with an Intel Xeon Processor E5-1650 processor (3.5GHz) and 32GB RAM. 
To fairly compare the efficiency of the index structures in terms of node quantity and size, we tune the size of each index node to be approximately 1 KB. For each experiment, the reported measurements are averaged over 5 runs.

\subsection{Dataset}
\noindent

\rewrite{
We use a synthesized YCSB dataset and two real-world datasets, Wikipedia data dump and Ethereum transaction data, to conduct a thorough evaluation of SIRI.
}

\subsubsection{YCSB} \rewrite{We generate the key-value dataset using YCSB according to the parameters shown in Table~\ref{table:para}. The lengths of the keys range from 5 bytes to 15 bytes, while the values have an average length of 256 bytes. The total number of records varies from 10,000 to 2,560,000. The dataset contains three types of workloads, read, write and mixed workload with 50\% write operations. We use Zipfian distribution to simulate the scenarios where the data in the workload is skewed to different degrees, where the Zipfian parameter $\theta$ equals 0 represents all records have an equal possibility to be selected into the workload, while higher $\theta$ value means only a smaller range of the records have extremely high possibilities to be selected. We also generate the overlapped workloads to test the capability of deduplication with increasing similarity in the contents, as described in Section~\ref{subsubsec:storage-collaborative}.}

\subsubsection{WIKI} \rewrite{The wiki dataset is real-world Wikipedia data dumps\footnote{\url{https://dumps.wikimedia.org/enwiki/}} of the extracted page abstracts. The key of the dataset is the URL of the Wikipedia page, the length of which ranges from 31 bytes to 298 bytes and has an average of 50 bytes. While the value of the dataset is the extracted abstract in plain text format, the length of which ranges from 1 byte to 1036 bytes, having an average of 96 bytes. We collect 6 data dumps covering the data changes in three months and divide the data into 300 versions. Each version has an average size of 855MB. We generate the read and write workload using keys uniformly selected from the dataset to test the throughput.}

\subsubsection{Ethereum Transactions} \rewrite{We use real-world Ethereum transaction data\footnote{\url{https://cloud.google.com/blog/products/data-analytics/ethereum-bigquery-public-dataset-smart-contract-analytics}} from Block 8900000 to 9200000, where the key is the 64-bytes block hash and the value is the RLP (Recursive Length Prefix) encoded raw transaction data. The length of the raw transaction ranges from 100 bytes to 57738 bytes with an average of 532 bytes. RLP is the main encoding method used to serialize objects in Ethereum, which is also used to encode raw transactions. In Ethereum, each block naturally makes a new version. }

\begin{figure*}[ht]
\centering
\subfigure[$\theta$ = 0, write ratio = 0] {
  \label{fig:exp:tps:a}
  \includegraphics[width=0.25\textwidth]{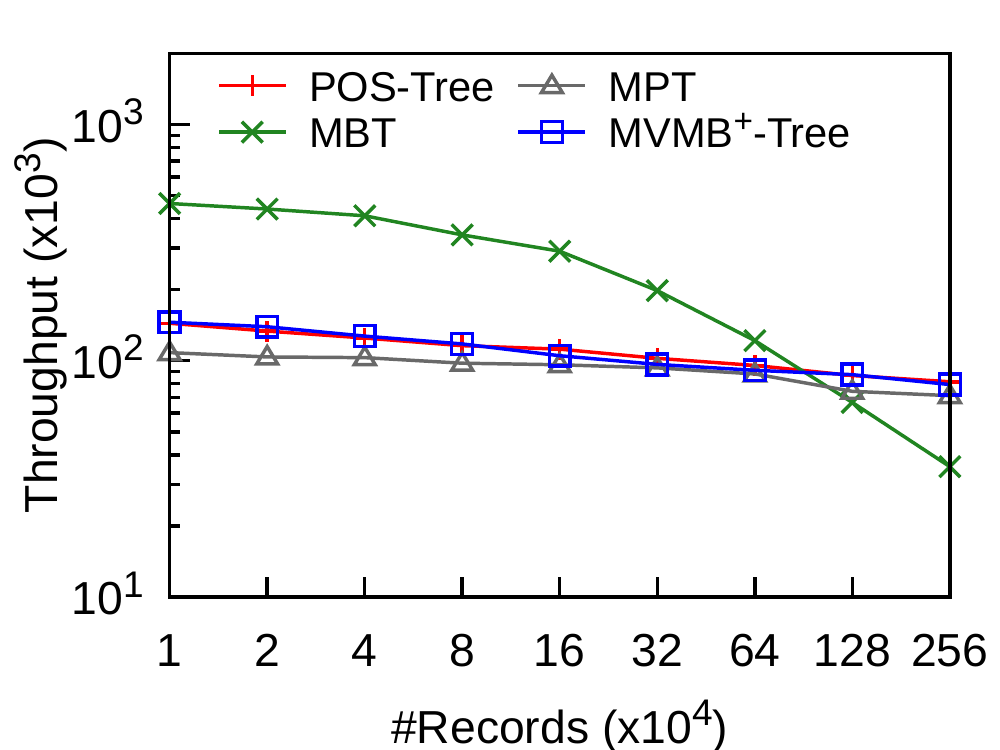}
}
\subfigure[$\theta$ = 0, write ratio = 0.5] {
  \label{fig:exp:tps:b}
  \includegraphics[width=0.25\textwidth]{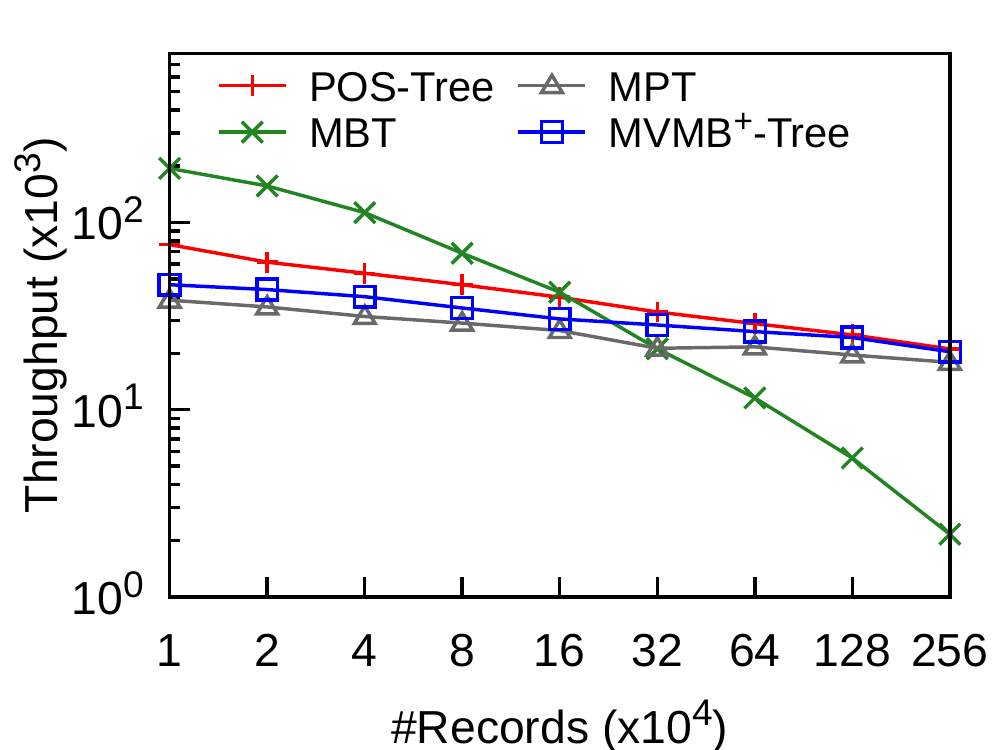}
}
\subfigure[$\theta$ = 0, write ratio = 1] {
  \label{fig:exp:tps:c}
  \includegraphics[width=0.25\textwidth]{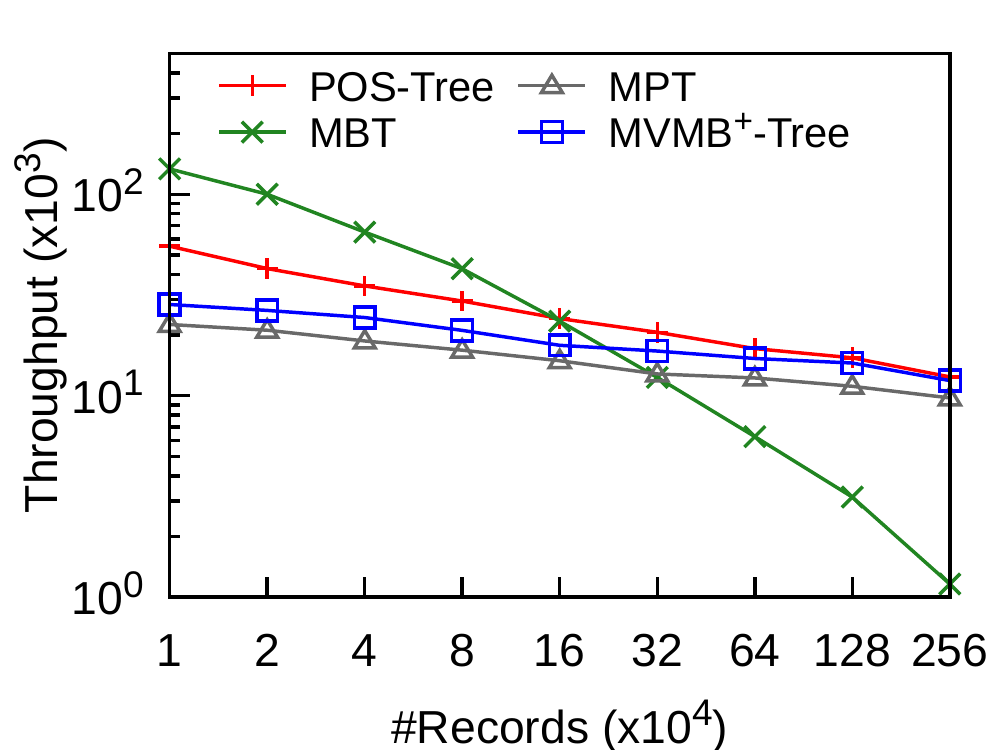}
}
\subfigure[$\theta$ = 0.5, write ratio = 0] {
  \label{fig:exp:tps:d}
  \includegraphics[width=0.25\textwidth]{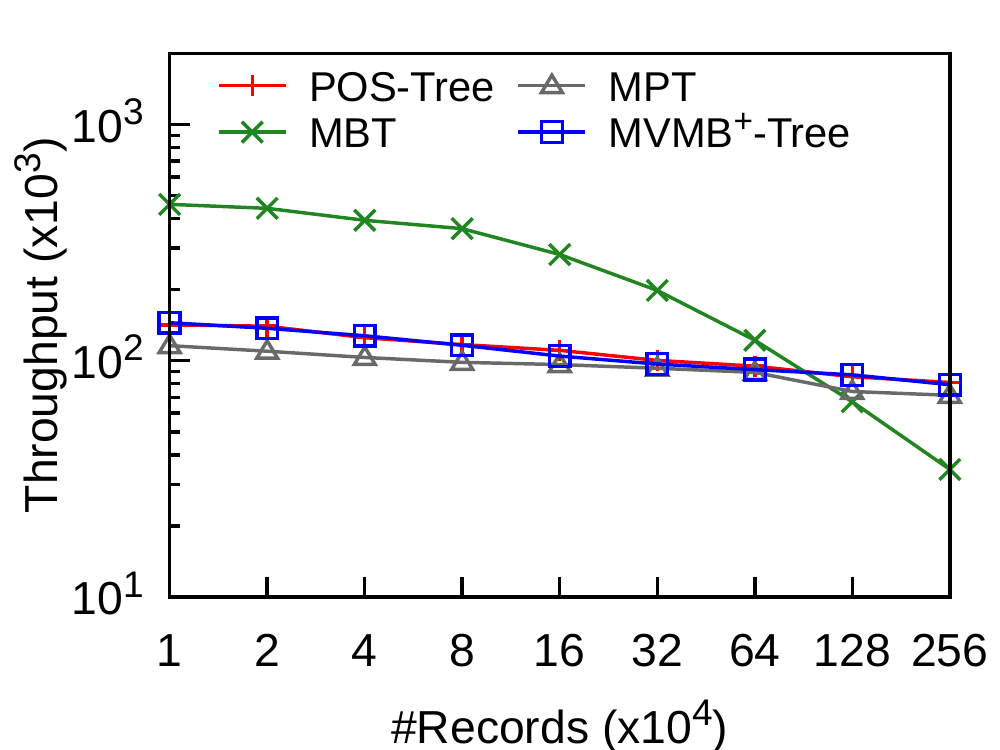}
}
\subfigure[$\theta$ = 0.5, write ratio = 0.5] {
  \label{fig:exp:tps:e}
  \includegraphics[width=0.25\textwidth]{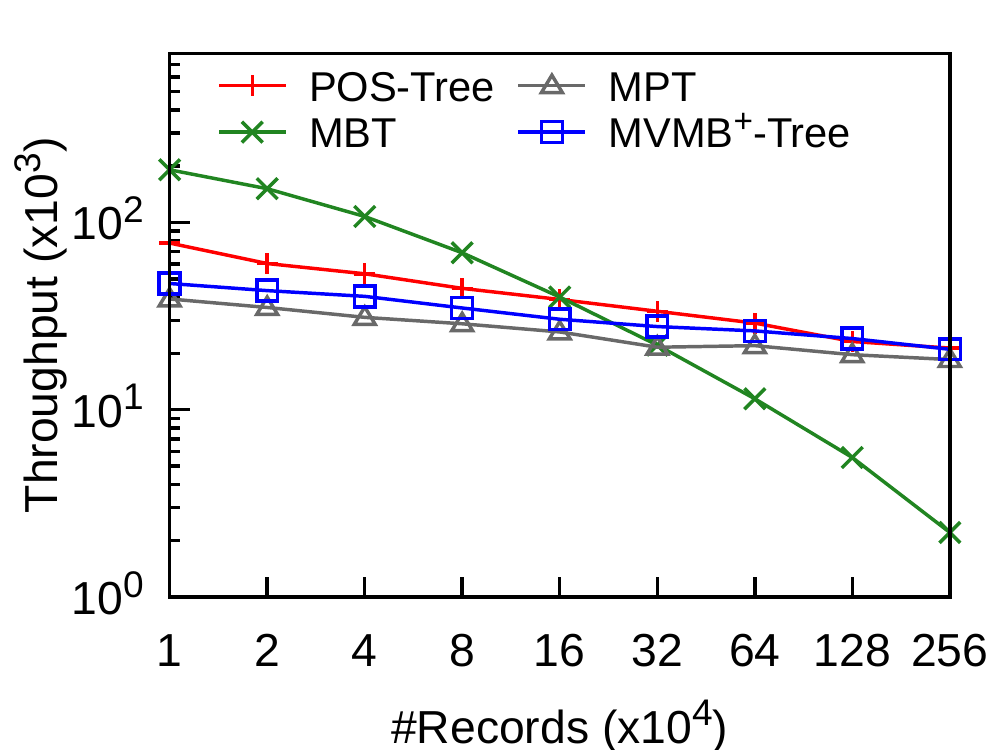}
}
\subfigure[$\theta$ = 0.5, write ratio = 1] {
  \label{fig:exp:tps:f}
  \includegraphics[width=0.25\textwidth]{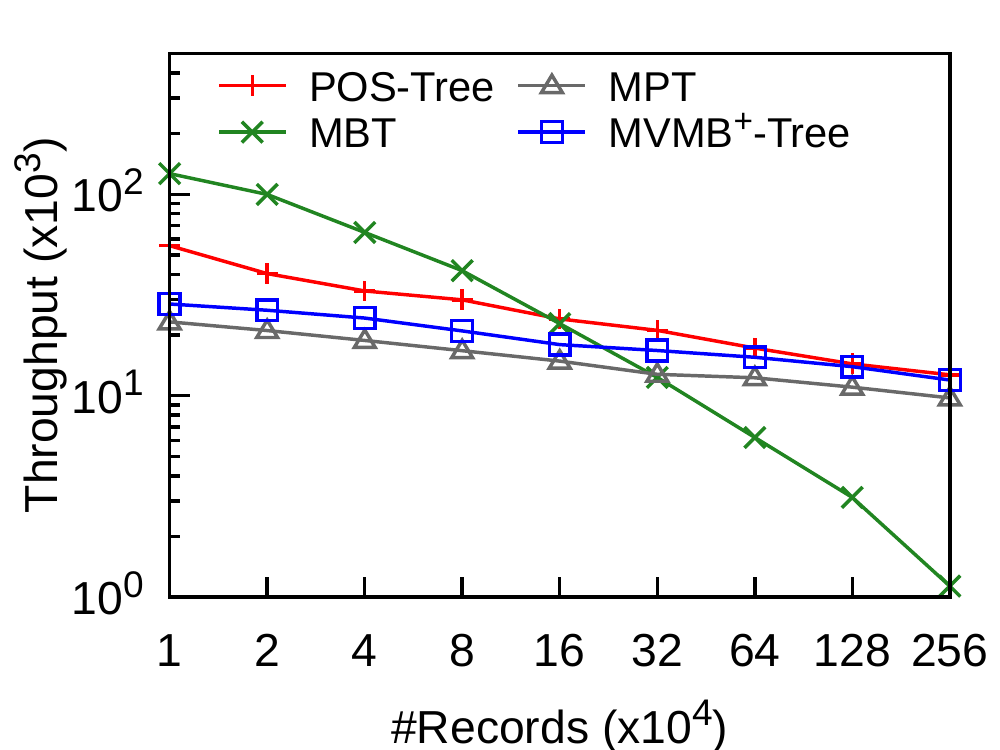}
}
\subfigure[$\theta$ = 0.9, write ratio = 0] {
  \label{fig:exp:tps:g}
  \includegraphics[width=0.25\textwidth]{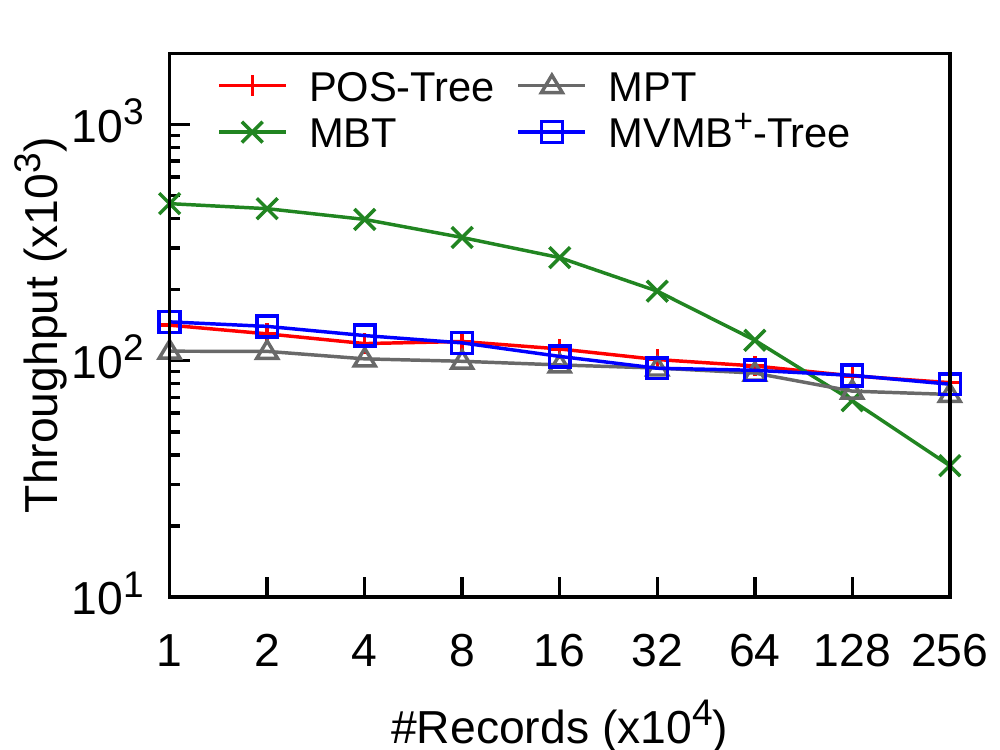}
}
\subfigure[$\theta$ = 0.9, write ratio = 0.5] {
  \label{fig:exp:tps:h}
  \includegraphics[width=0.25\textwidth]{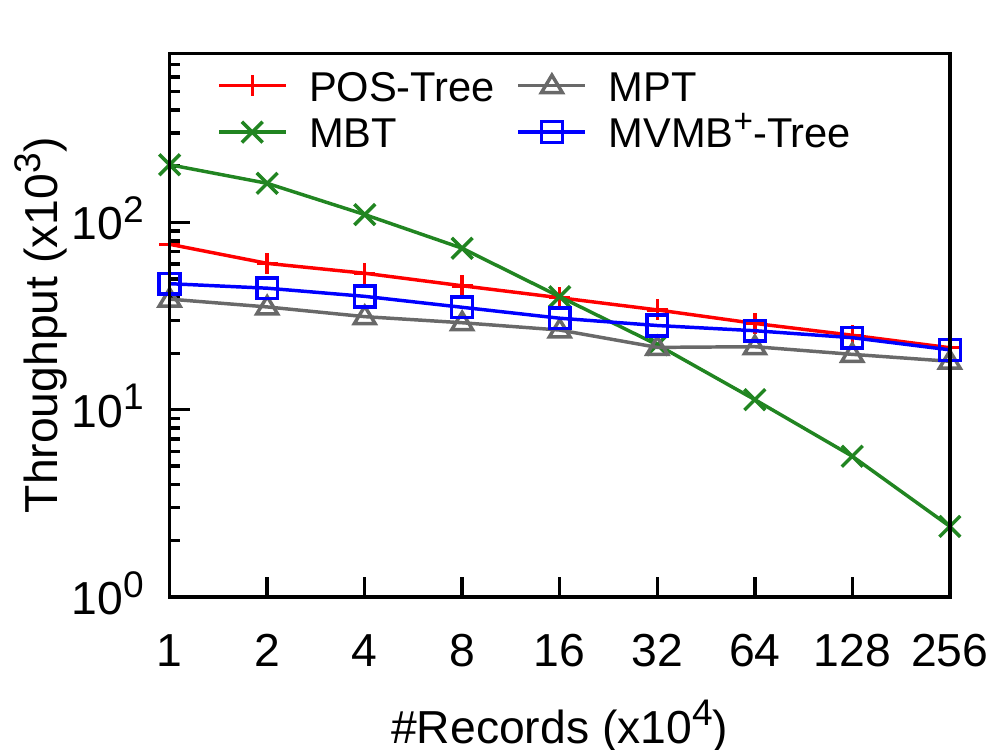}
}
\subfigure[$\theta$ = 0.9, write ratio = 1] {
  \label{fig:exp:tps:i}
  \includegraphics[width=0.25\textwidth]{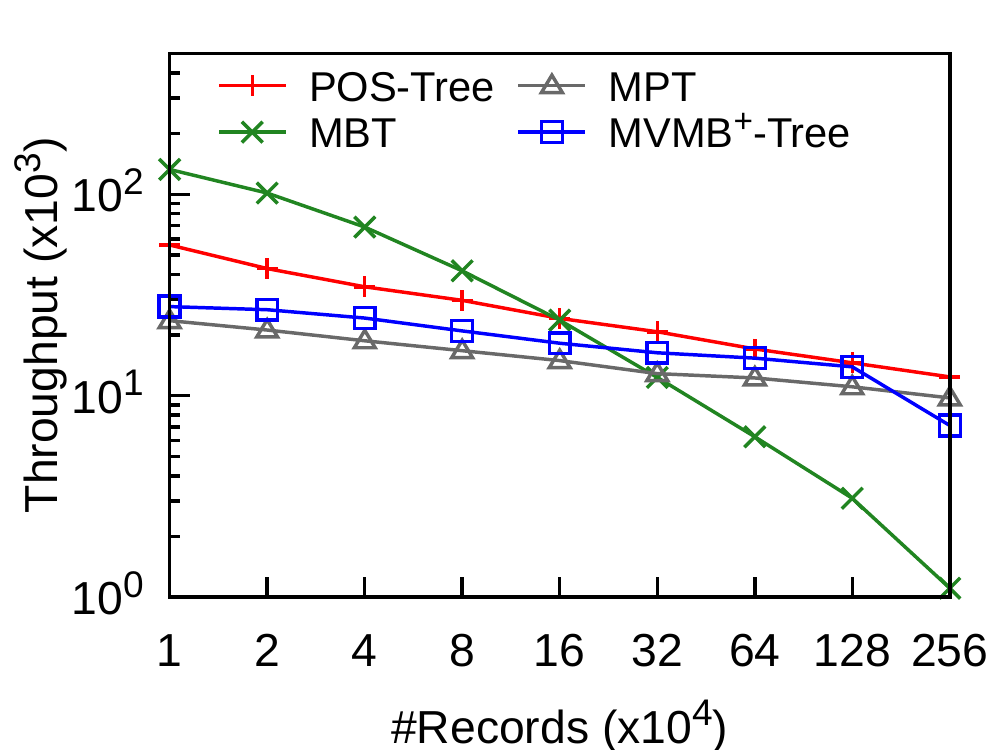}
}
\caption{Throughput on YCSB}
\label{fig:exp:tps}
\end{figure*}

\begin{figure*}
  \centering
  \begin{minipage}{0.5\textwidth}
    \vspace{-5pt}
    \subfigure[Wiki]{    
      \centering
      \includegraphics[width=0.47\textwidth]{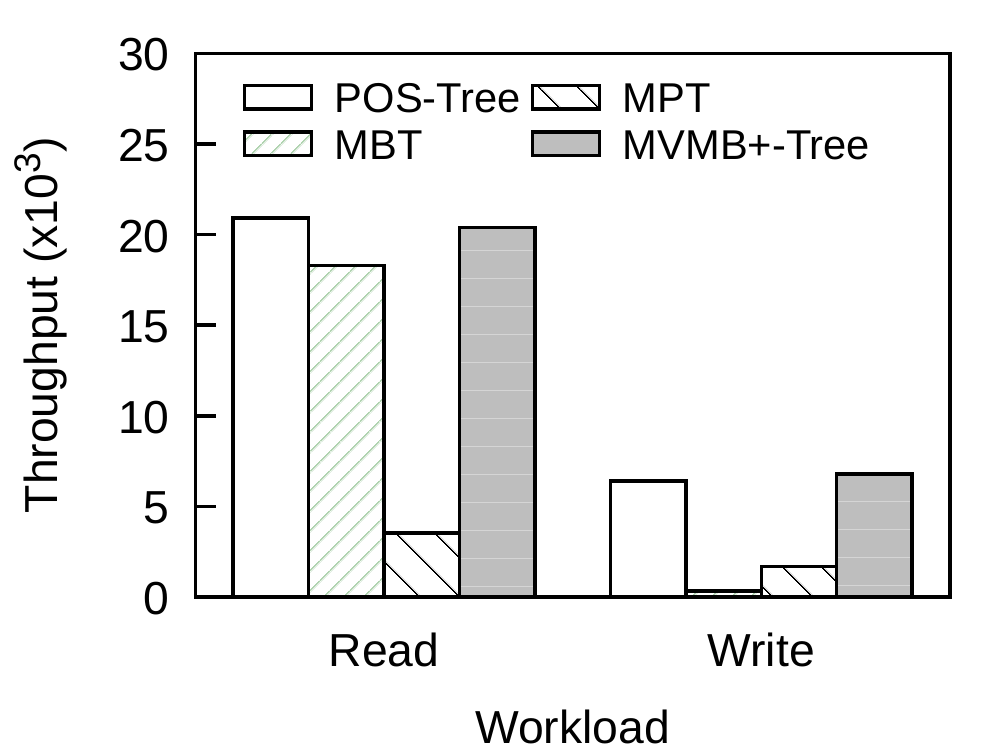}
      \label{fig:exp:wiki_tps}
    }
    \subfigure[Ethereum Transaction]{
      \centering
      \includegraphics[width=0.47\textwidth]{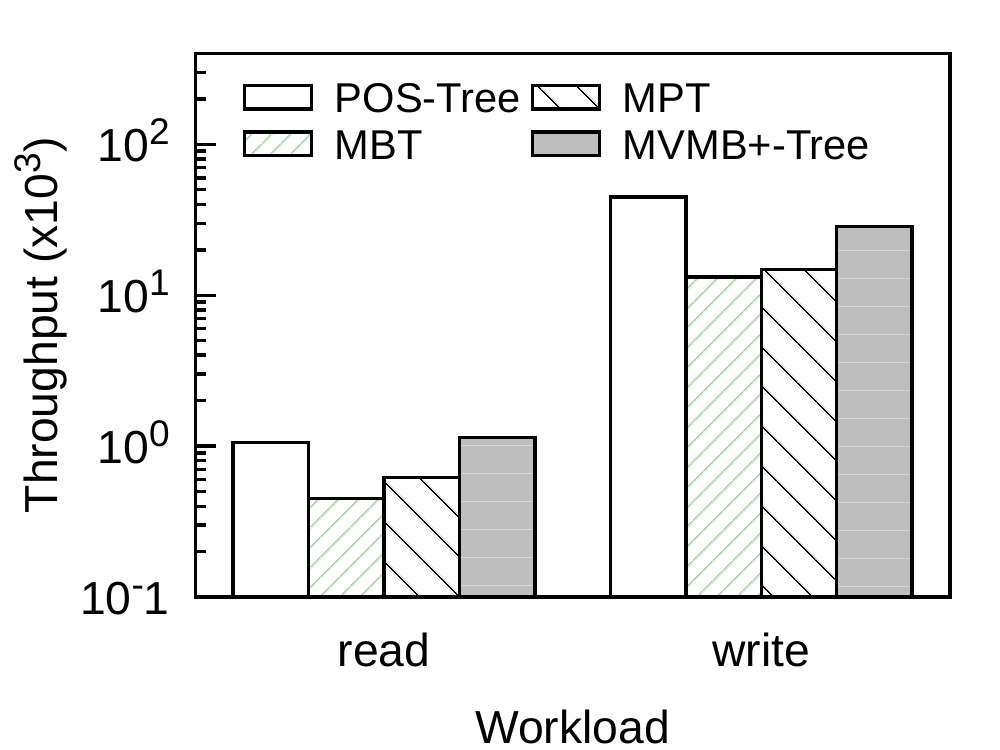}
      \label{fig:exp:eth_tps}
    }
    \vspace{-15pt}
    \caption{Throughput on real world datasets}
  \end{minipage}
  \begin{minipage}{0.24\textwidth}
    \subfigure{    
      \centering
      \includegraphics[width=0.96\textwidth]{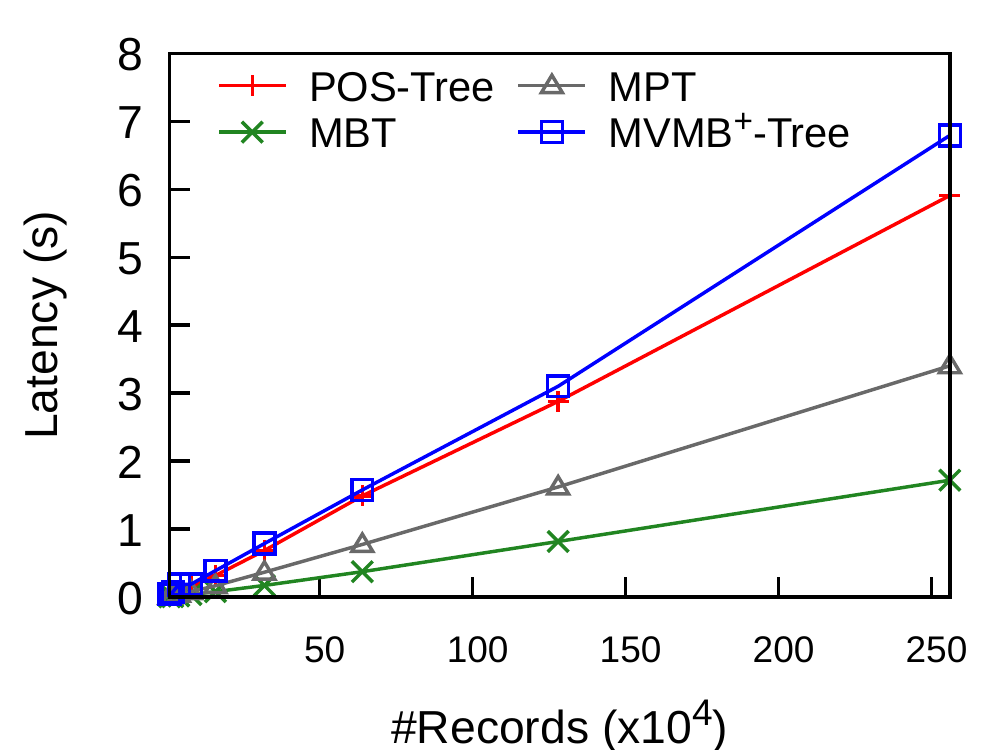}
      \label{fig:exp:diff}
    }
    \caption{Diff performance}
  \end{minipage}
  \begin{minipage}{0.24\textwidth}
    \subfigure{
      \centering
      \includegraphics[width=0.96\textwidth]{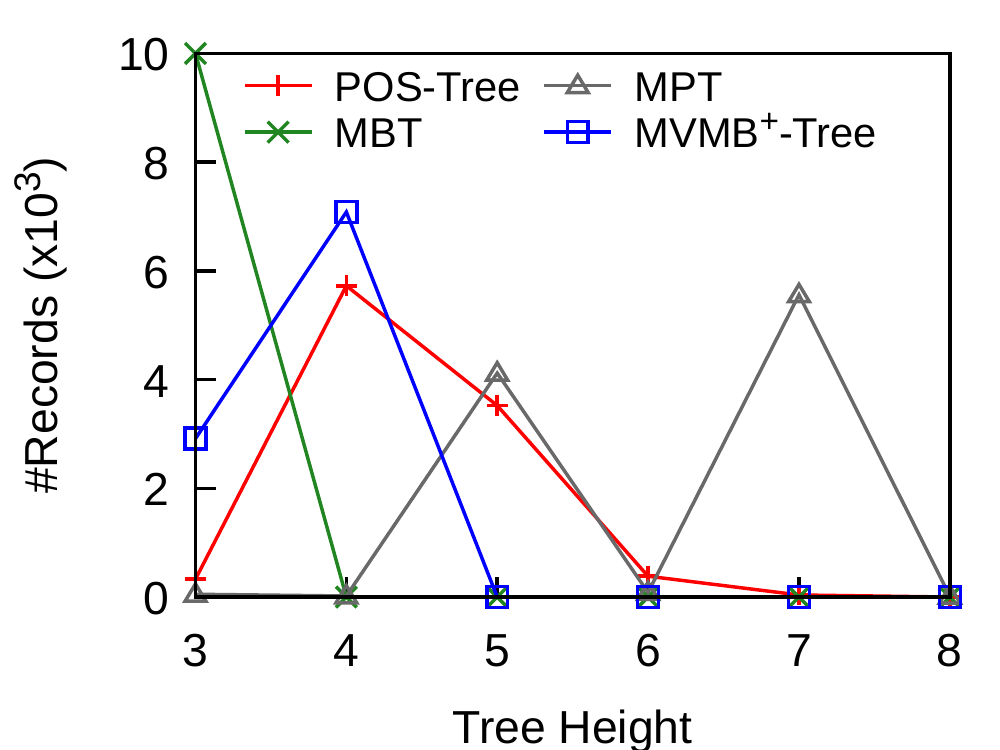}
      \label{fig:exp:height}
    }
    \caption{Tree height}
  \end{minipage}
\end{figure*}

\subsection{Implementation}
\noindent

In this section, we briefly describe the implementation of selected indexes and the baseline.
We port the Ethereum's implementation~\cite{web:ethereum} of MPT to our experiment environment, which adopts the path compaction optimization.
The implementation of MBT is based on the source code provided in Hyperledger Fabric 0.6~\cite{web:hyperledger}.
We further make it immutable and add index lookup logic, which is missing in the original implementation.
For \mytree, we use the implementation in Forkbase~\cite{wang:2018}.
Moreover, we further apply batching techniques, taking advantage of the bottom-up build order, to reduce the number of tree traversal and hash calculations significantly.
Lastly, to compare \myindex and non-\myindex structures, we implement an immutable B$^+$-tree with tamper evidence support, called Multi-Version Merkle B$^+$-tree (MVMB$^+$-Tree), as the baseline.
We replace the pointers stored in index nodes with the hash of their immediate children and maintain an additional table from the hash to the actual address.
For all the structures, we adopt node-level copy-on-write to achieve the data immutability.

\subsection{Throughput and Latency}
\noindent

We evaluate the three candidates and the baseline from a traditional view in this part, where throughput and latency are the major measurements.

\vspace{10pt}
\subsubsection{Throughput}

\noindent

\begin{figure*}
  \centering
  \begin{minipage}{0.4\textwidth}
    \centering
    \includegraphics[scale=0.4]{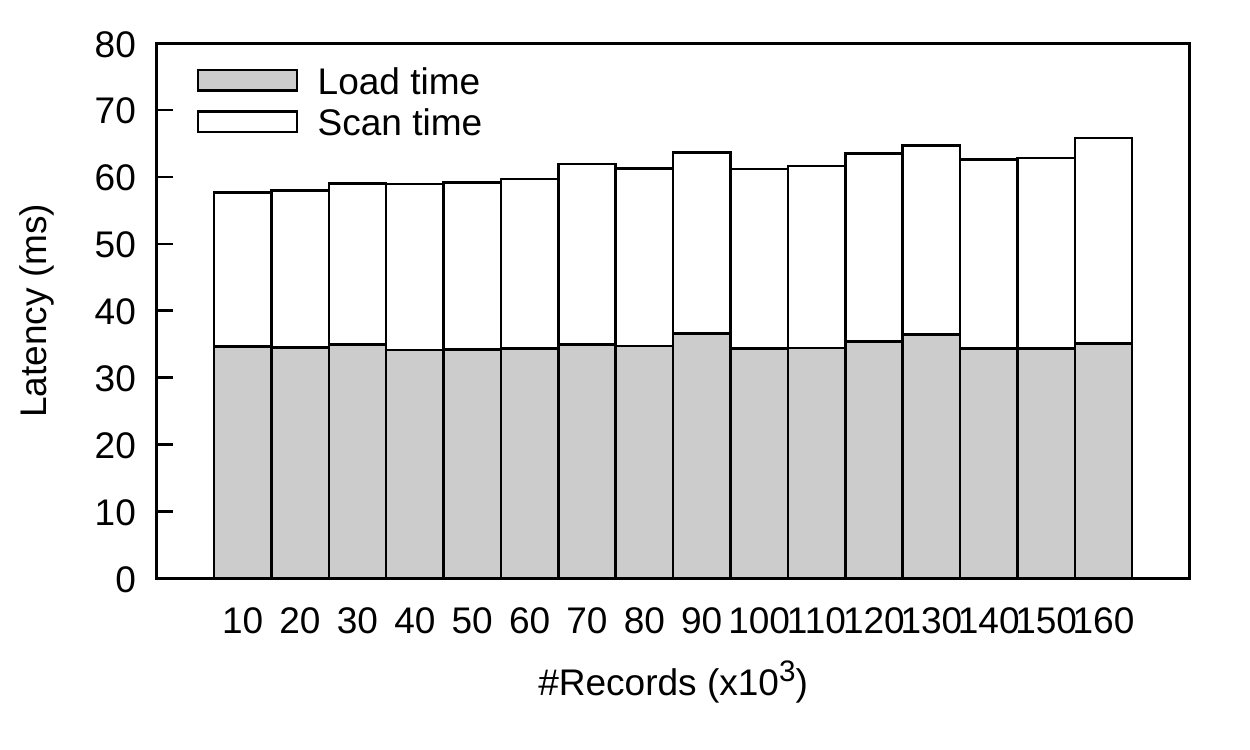}
    \vspace{1ex}
    \caption{MBT breakdown latency}
    \label{fig:exp:mbt}
  \end{minipage}
  \begin{minipage}{0.5\textwidth}
    \vspace{-5pt}
    \centering
    \subfigure[Storage]{    
      \centering
      \includegraphics[width=0.47\textwidth]{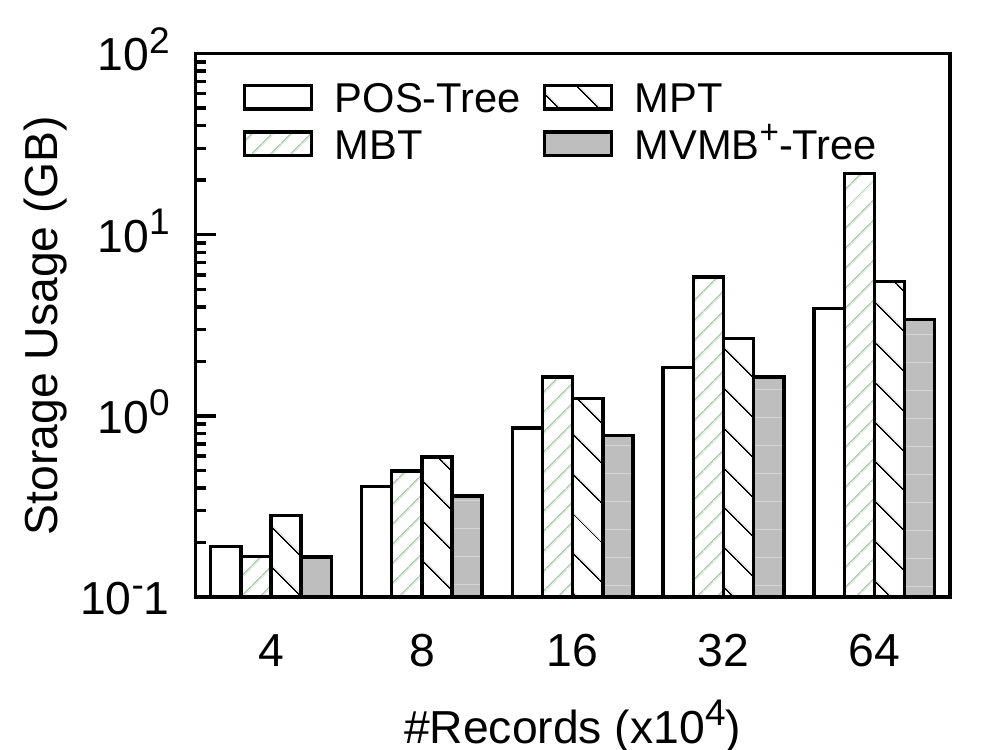}
      \label{fig:exp:basic_sto}
    }
    \subfigure[Number of nodes]{
      \centering
      \includegraphics[width=0.47\textwidth]{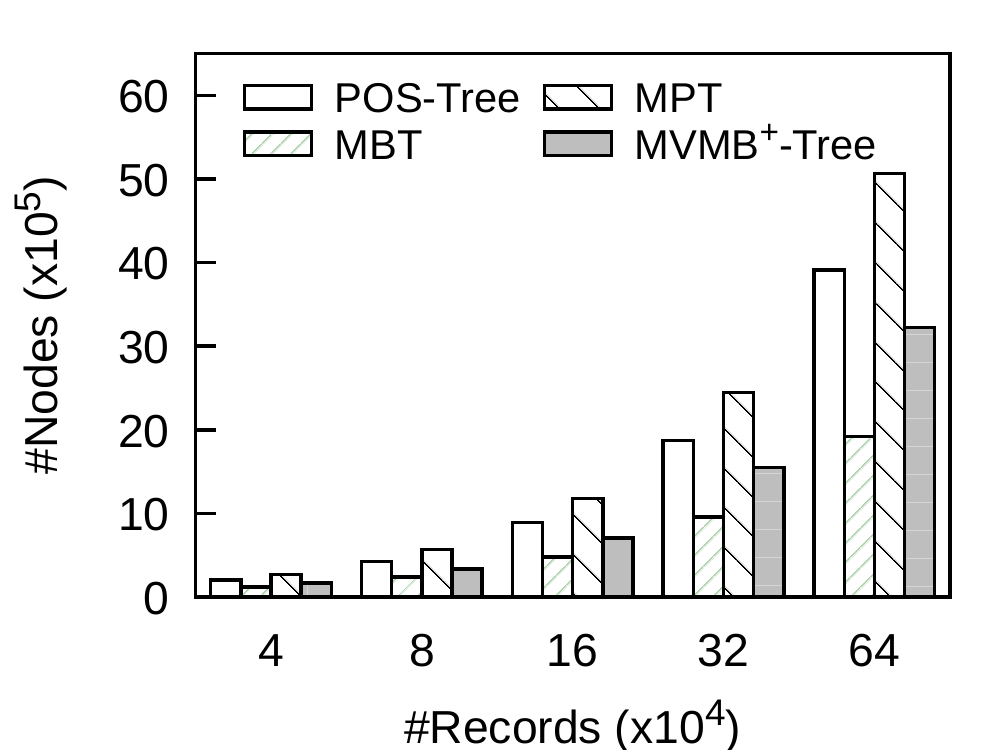}
      \label{fig:exp:basic_chk}
    }
    \vspace{-15pt}
    \caption{Performance on single group data access}
    \label{fig:exp:basic}
  \end{minipage}
\end{figure*}

First, we evaluate the throughput using the YCSB dataset.
We run the read, the write and the mixed workloads under diverse data size and skewness.
The results are illustrated in Figure~\ref{fig:exp:tps}.
It can be observed that the throughput of all indexes decreases as the number of data grows and complies with the operation bound formulated in Section~\ref{subsec:operation_bounds}.
Figure~\ref{fig:exp:tps:a} shows the throughput for the read workload with uniform data.
The throughput of \mytree is 0.95x - 1.06x of the baseline while MPT is only 0.74x - 0.96x of the baseline. The throughput of MBT drops quickly from 3.2x to 0.45 of the baseline due to the dominating leaf loading and scanning process. 
As it is shown in Figure~\ref{fig:exp:mbt}, the time to traverse the tree and load the nodes keeps constant, while time to scan leaf node keeps increasing.
For the write workload shown in Figure~\ref{fig:exp:tps:c}, we can observe a similar trend. However, \mytree performs 1.04x-1.9x better than the baseline taking advantage of the batching techniques and the bottom-up building process.


\begin{figure*}
  \centering
  \subfigure[Read Balanced]{    
    \centering
    \includegraphics[scale=0.4]{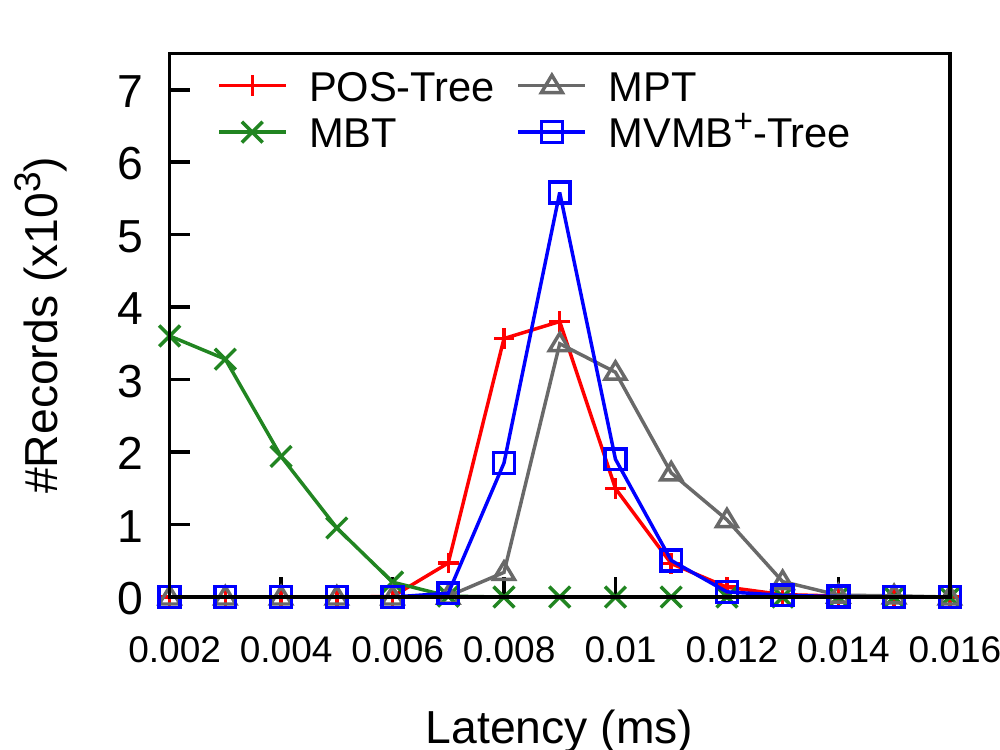}
    \label{fig:exp:lat:a}
  }
  \subfigure[Read Skewed]{
    \centering
    \includegraphics[scale=0.4]{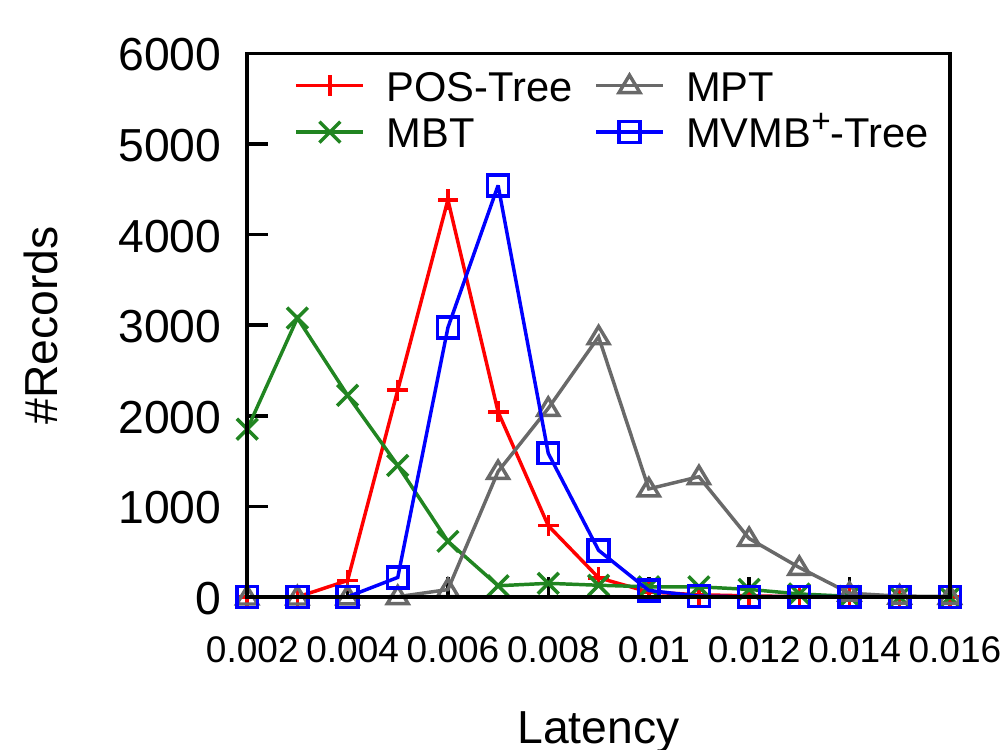}
    \label{fig:exp:lat:b}
  }
  \subfigure[Write Balanced]{
    \centering
    \includegraphics[scale=0.4]{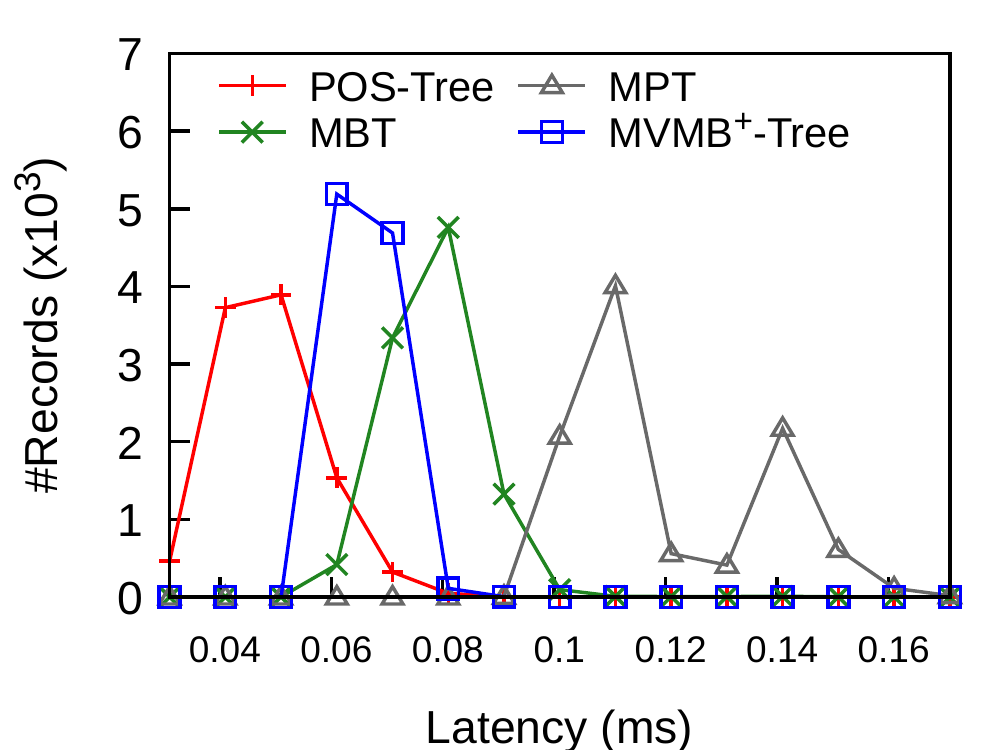}
    \label{fig:exp:lat:c}
  }
  \subfigure[Write Skewed]{
    \centering
    \includegraphics[scale=0.4]{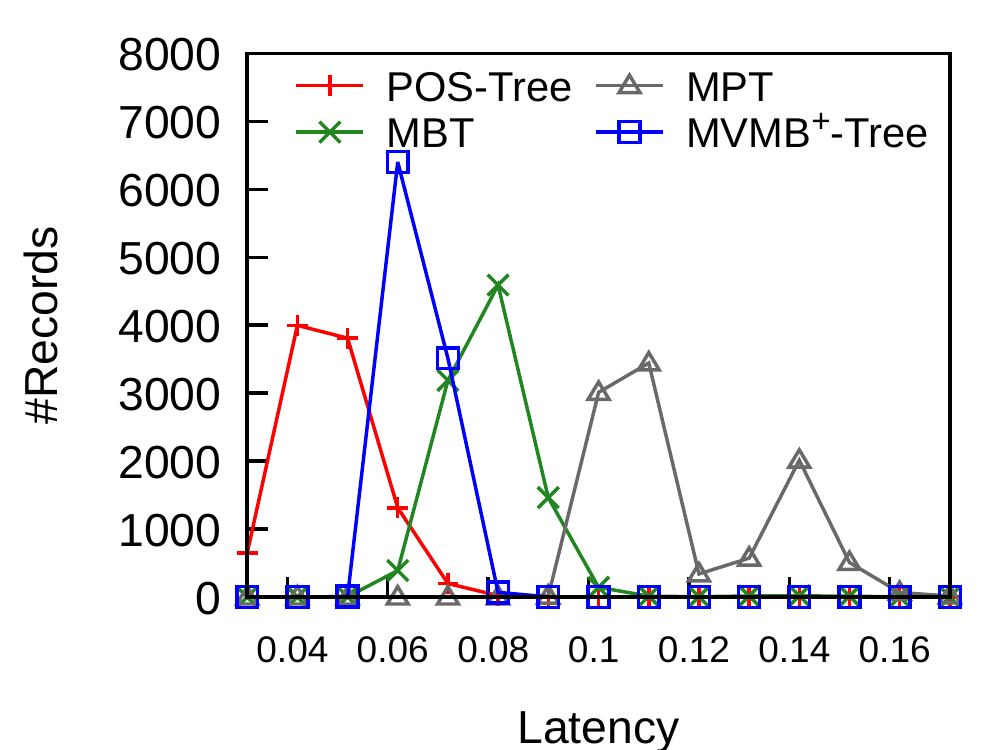}
    \label{fig:exp:lat:d}
  }
  \caption{Latency on YCSB}
  \label{fig:exp:lat}
\end{figure*}

\begin{figure*}
  \centering
  \begin{minipage}{.49\textwidth}
    \subfigure[Read]{    
      \centering
      \includegraphics[width=0.47\textwidth]{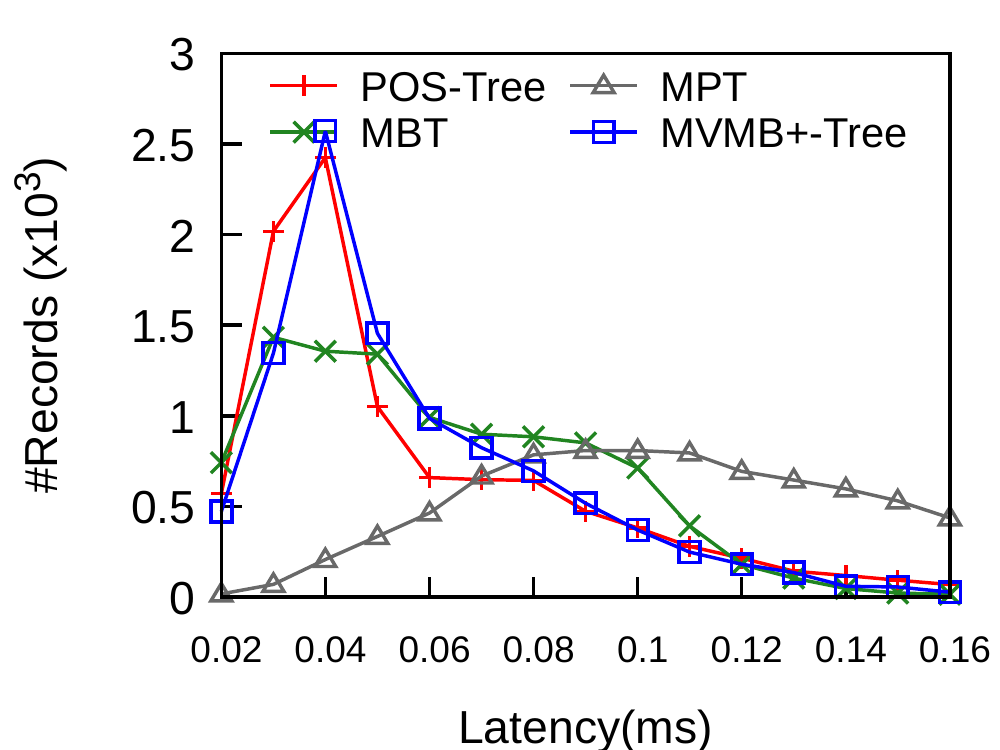}
    }
    \subfigure[Write]{
      \centering
      \includegraphics[width=0.47\textwidth]{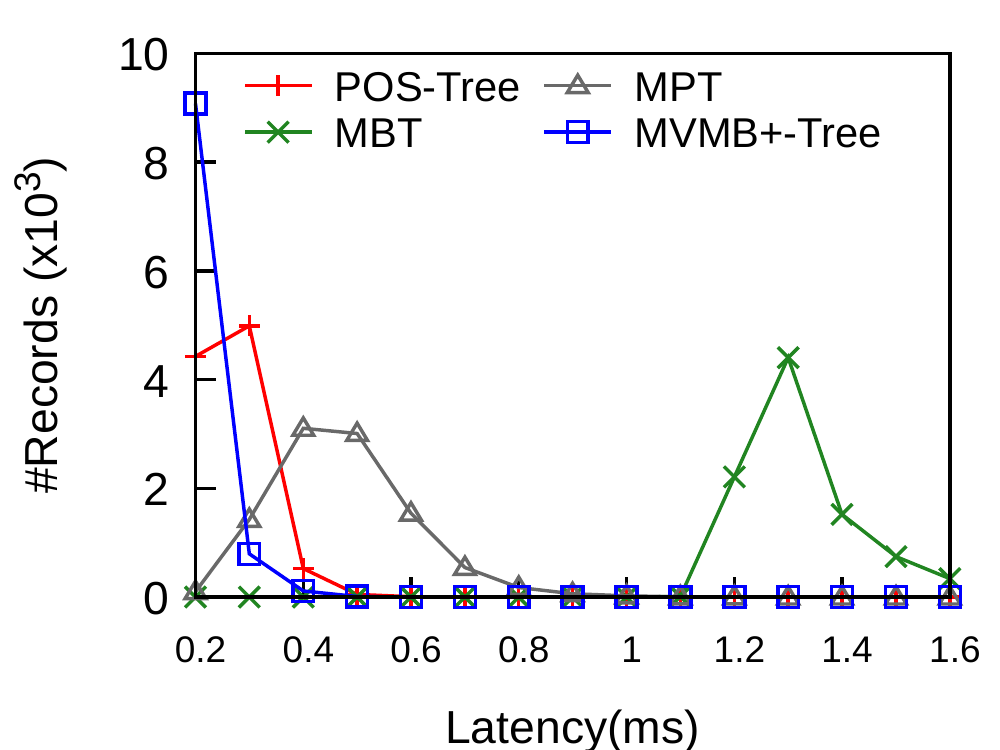}
    }
    \vspace{-10pt}
    \caption{Latency on Wiki data}
    \label{fig:exp:wiki_lat}
  \end{minipage}
  \begin{minipage}{.49\textwidth}
    \subfigure[Read]{
      \centering
      \includegraphics[width=0.47\textwidth]{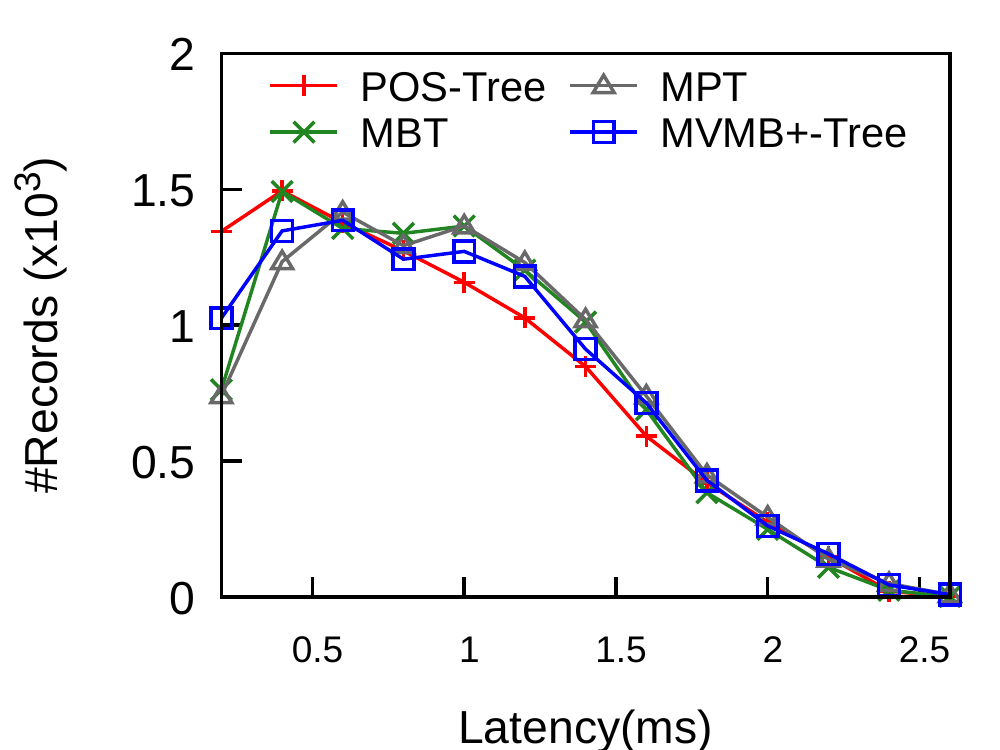}
    }
    \subfigure[Write]{
      \centering
      \includegraphics[width=0.47\textwidth]{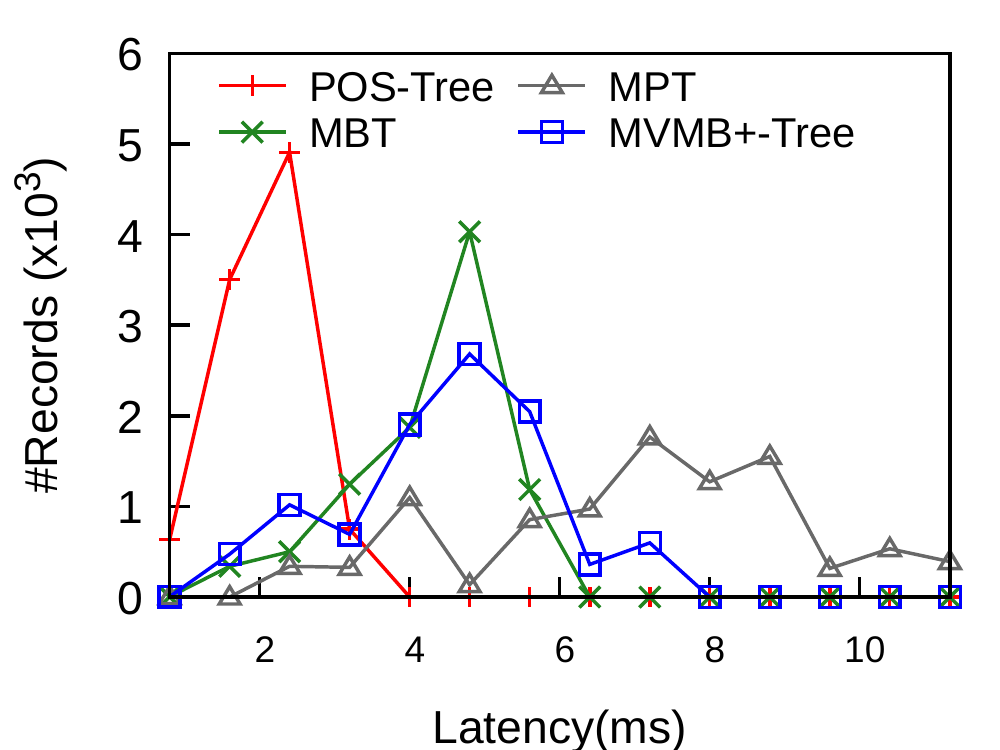}
    }
    \vspace{-10pt}
    \caption{Latency on Ethereum transaction data}
    \label{fig:exp:eth_lat}
  \end{minipage}
\end{figure*}

By comparing Figure~\ref{fig:exp:tps} horizontally, we can observe that the throughput of all data structures decreases drastically as the ratio of write operations increases.
This is due to the cost of node creation, memory copy and cryptographic function computation.
The absolute throughput drops over 6.6x in the comparison of the largest dataset for \mytree and baseline while it drops 30x for MBT and 7.3x for MPT.
By comparing Figure~\ref{fig:exp:tps} vertically, we can observe that there is no change in throughput for all index structures when $\theta$ changes from 0 to 0.9.
Therefore, we can conclude that they are all resilient to data skewness.

It is also worth noting that, compared with MBT, the other three structures perform much more steady for both read and write workloads.
Meanwhile, \mytree outperforms MPT in all cases and has comparable performance compared to our baseline index.

\rewrite{Next, we run the experiment on the Wiki dataset. The system first load the entire dataset batched in 300 versions, and then execute the read and write workloads that is uniformly selected. Figure~\ref{fig:exp:wiki_tps} demonstrates the results that are aligned with those in the YCSB experiment.}

\rewrite{
Lastly, for the experiments on Ethereum data, we simulate the way blockchain stores the transactions.
For each block, we build an index on transaction hash for all transactions within that block and store the root hash of the tree in a global linked list.
Versions are naturally created at a block granularity.
For write operations, the system appends the new block of transactions to the global linked list while for lookup operations, it scans the linked list for the block containing the transaction, and traverses the index to obtain the value.
Figure~\ref{fig:exp:eth_tps} shows the result of this experiment.
It can be observed that
\mytree outperforms other indexes in write workloads.
This is because we are building indexes for each block instead of a global index.
Further,
instead of insert/update operations, we perform batch loading from scratch.
In this case, \mytree's bottom-up building process is superior to the MPT's and MVMB$^+$-Tree's top-down building process, as it only traverses the tree and creates each node once.
Another difference is that the throughput of read workload is lower than that of  the write workload mainly due to the additional block scanning time.}

\subsubsection{Latency and Path Length}

\noindent

In this experiment, we measure the latency of each read and write operation and calculate the distribution with balanced and skewed data.
For the YCSB dataset, read-only and write-only workloads are fed into the indexes with balanced ($\theta=0$) and highly skewed ($\theta=0.9)$ distributions.
The dataset used in this test contains 160,000 keys.
We run 10,000 operations and pictured the latency distribution in Figure~\ref{fig:exp:lat}.
The x-axis is the range of the latency and the y-axis is the number of records fell in that latency range.
It can be seen from the figure that the rankings among the indexes coincide with the previous conclusion -- \mytree performs the best for both read and write workloads while MPT performs the worst.
Meanwhile, MPT has several peak points, representing operations accessing data stored in different levels of the tree.
MBT experiences the most dramatic changes between read and write workloads.
It outperforms all the candidates in the read workloads but is worse than \mytree in write workload.

To take a closer observation of how the workloads affect the candidates, we further gather the traversed tree height of each operation for the write-only workload with the uniform distribution.
The results are shown in Figure~\ref{fig:exp:height}, where the x-axis represents the height of the lookup path and the y-axis indicates the number of operations.
Most operations have to visit 4-level nodes to reach the bottom-most level of~\mytree whilst 5- or 7-level nodes are frequently traversed for MPT.
The efficiency in MBT is also verified in the figure since all requests only need 3 levels to reach the bottom of the structure in both balanced and skewed scenarios.

\begin{figure*}
  \centering
  \begin{minipage}{.49\textwidth}
    \subfigure[Storage]{
      \centering
      \includegraphics[scale=0.40]{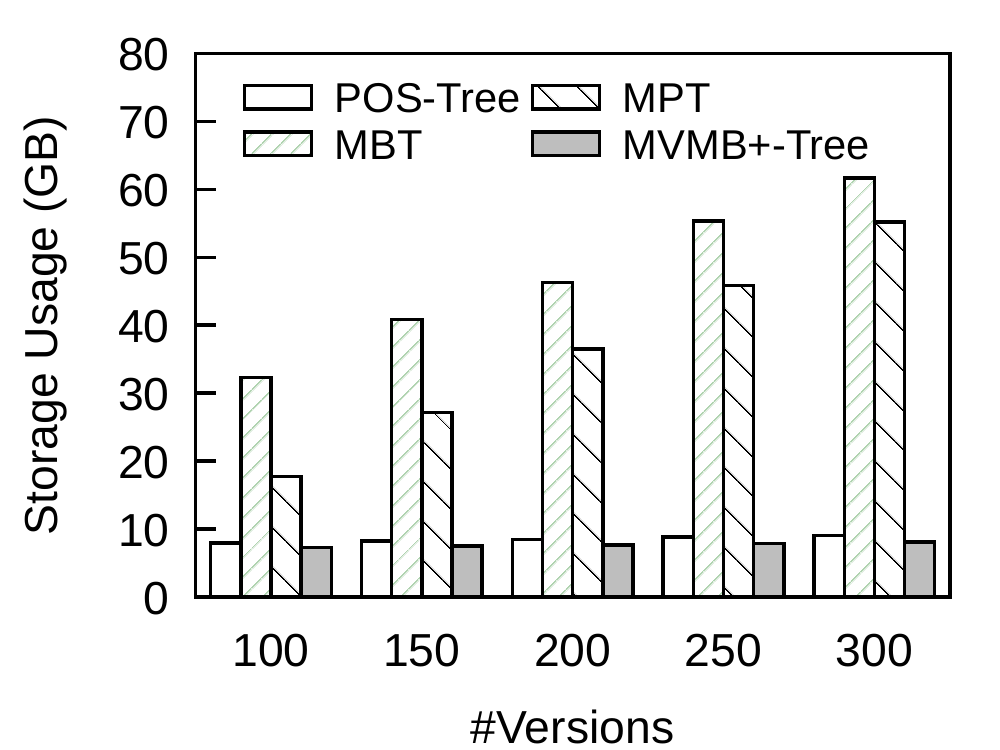}
    }
    \subfigure[Number of nodes]{    
      \centering
      \includegraphics[scale=0.40]{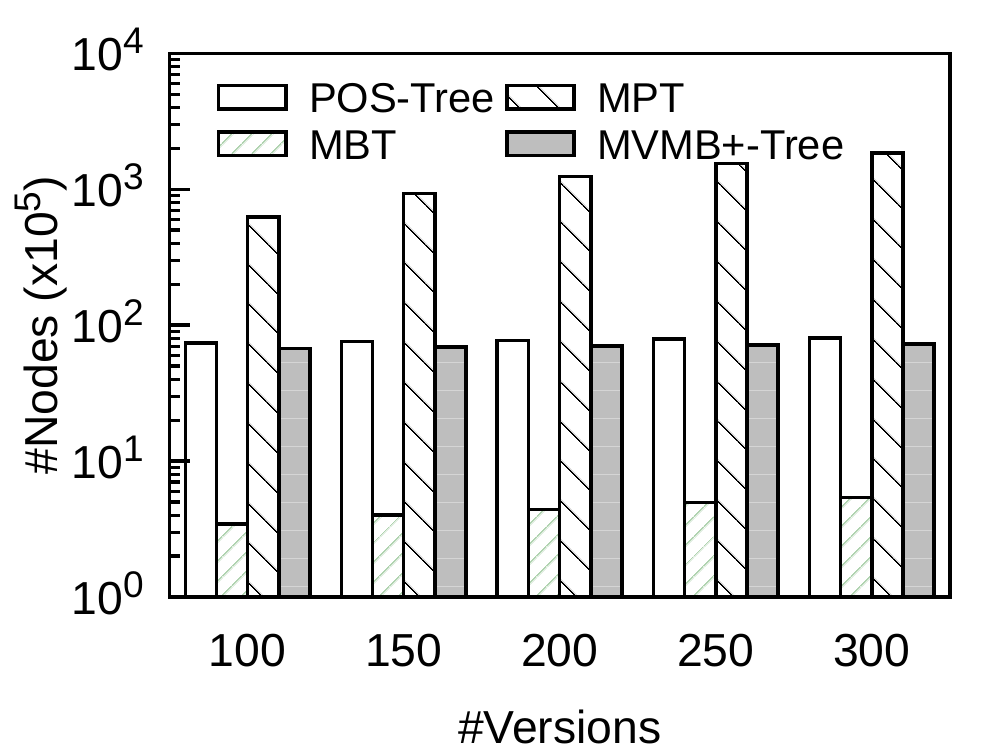}
    }
    \caption{Storage on Wiki data}
    \label{fig:exp:wiki_sto}
  \end{minipage}
  \begin{minipage}{.49\textwidth}
    \subfigure[Storage]{
      \centering
      \includegraphics[scale=0.40]{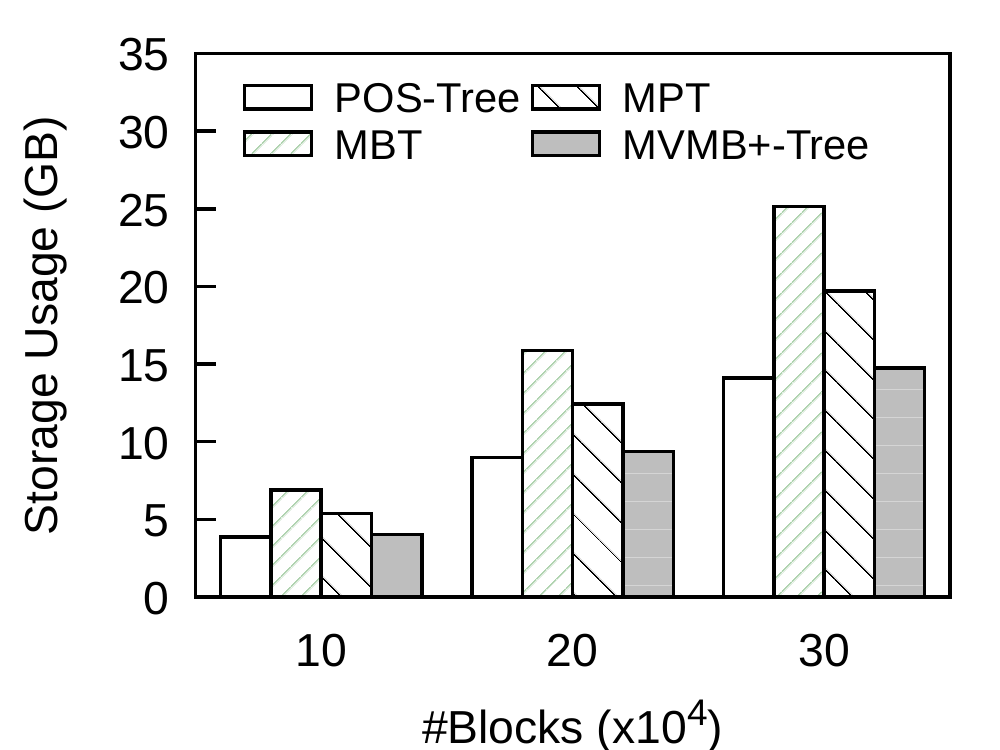}
    }
    \subfigure[Number of nodes]{    
      \centering
      \includegraphics[scale=0.40]{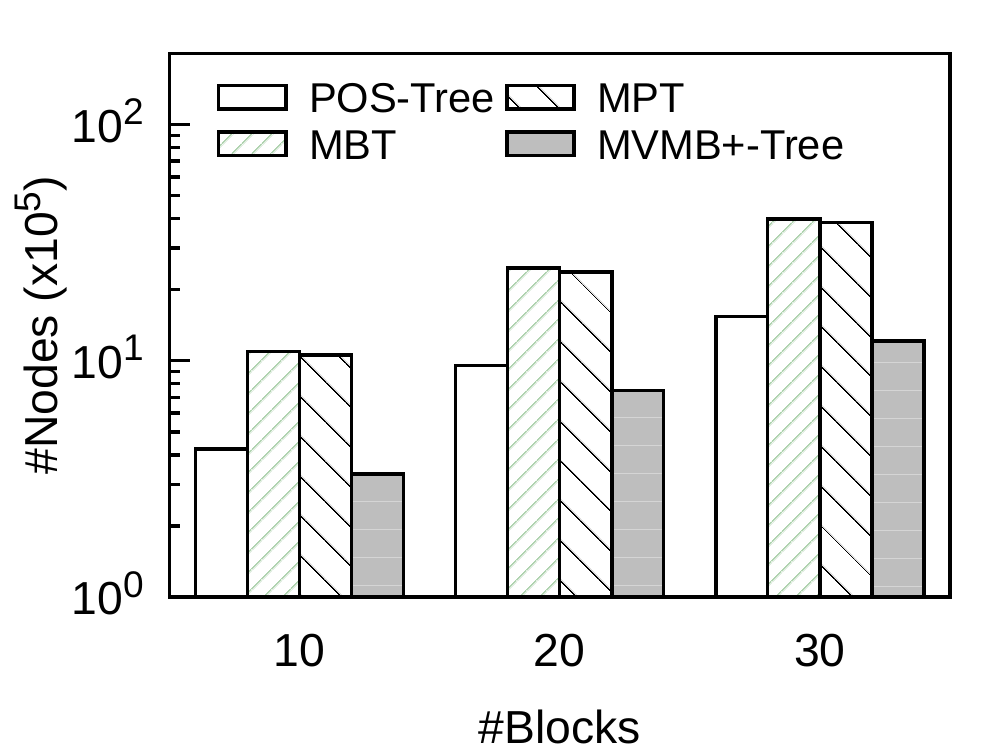}
    }
    \caption{Storage on Ethereum transaction data}
    \label{fig:exp:eth_sto}
  \end{minipage}
\end{figure*}

\rewrite{
We can obtain similar results and conclusions on the Wiki dataset as shown in Figure~\ref{fig:exp:wiki_lat}.
However, the experiment on Ethereum transaction data exhibits different trends as depicted in Figure~\ref{fig:exp:eth_lat}.
As can be observed, all the structures have similar read latency, caused by the dominant block scanning process.
}

We also run a diff workload to evaluate the performance of ``diff'' operations.
In the experiment, each structure loads two versions of data in random order.
A diff operation is performed between the two versions and the execution time is taken, as depicted in Figure~\ref{fig:exp:diff}.
All the candidates outperform the baseline due to the structurally invariant property.
Among which, MBT performs the best (4x of baseline) since the position of the nodes containing a specific data is static among all versions.
The logic of diff operation is the simplest, i.e., comparing the hash of the nodes at the corresponding position.
MPT performs 2x better than the baseline and 1.7x better than \mytree due to the simplicity that keys with the same length always lie in the same level of the tree.

\subsection{Storage}
\noindent

In this section, we evaluate the space consumption of the index structures under different use cases.

\subsubsection{Single Group Data Access}

\noindent

We first start with a simple case, where a dataset is accessed by multiple users.
There is no sharing of data or cross-department collaborative editing in this setting.
Therefore, the deduplication benefit is limited using \myindex.
In reality, such case often happens in-house within a single group of users from the same department.
Figure~\ref{fig:exp:basic_sto} shows the storage under different data sizes for the YCSB dataset.
There are two main factors affecting the space efficiency, i.e., the size of the node and the height of the tree.
On the one hand, larger tree height results in more node creations for write operations, which also increases the space consumption.
As an example, MPT performs badly since it has the largest tree height in our experiment setting.
It consumes the storage up to 1.6x higher than the baseline and up to 1.4x larger than \mytree.
On the other hand, a large node size means that even minor changes to the node could trigger the creation of a new substantial node, which hence leads to larger space consumption.
As can be seen, MBT performs the worst due to the largest node size it has in the implementation.
It consumes up to 6.4x the space of that used by the baseline.
\mytree, compared to the baseline, also has a larger node size variance due to content-defined chunking, leading to a greater number of large nodes.


\begin{figure*}[ht]
  \centering
  \subfigure[Storage]{
    \centering
    \includegraphics[scale=0.40]{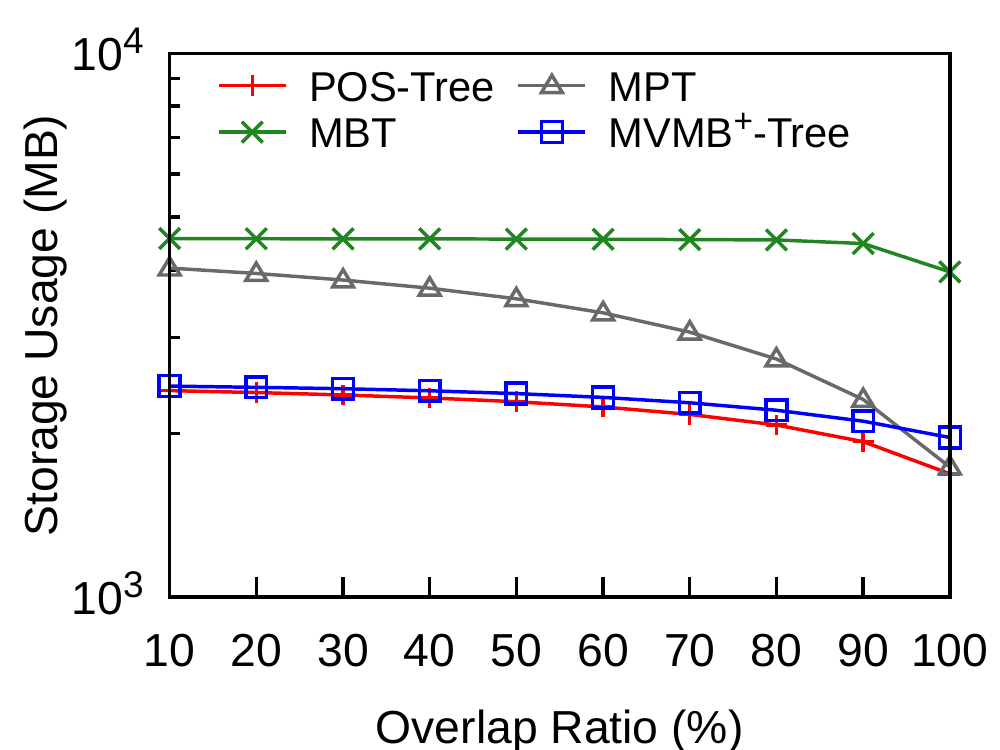}
    \label{fig:exp:overlap_sto}
  }
  \subfigure[Number of nodes]{
    \centering
    \includegraphics[scale=0.40]{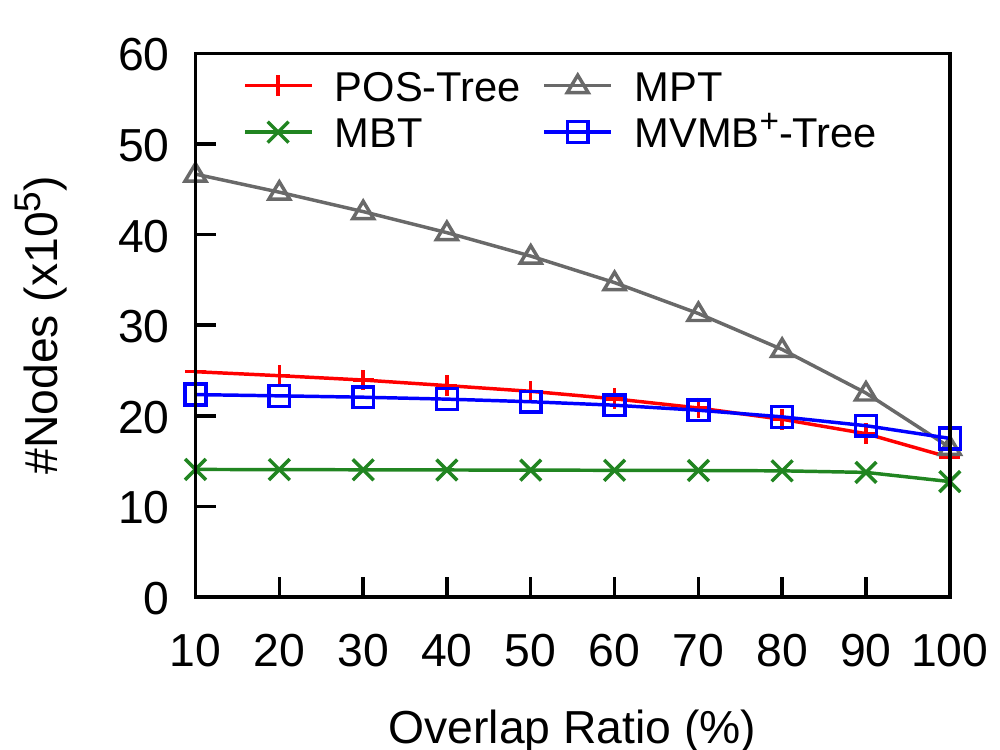}
    \label{fig:exp:overlap_chk}
  }
  \subfigure[Deduplication ratio]{
    \centering
    \includegraphics[scale=0.40]{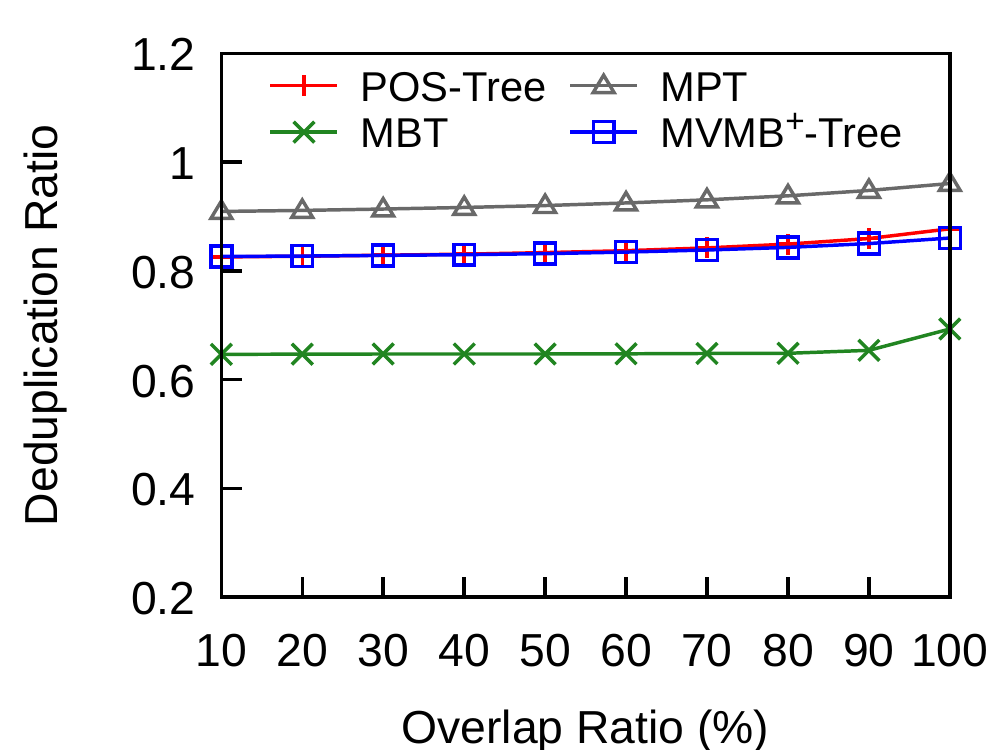}
    \label{fig:exp:overlap_dedup}
  }
  \subfigure[Node sharing ratio]{
    \centering
    \includegraphics[scale=0.40]{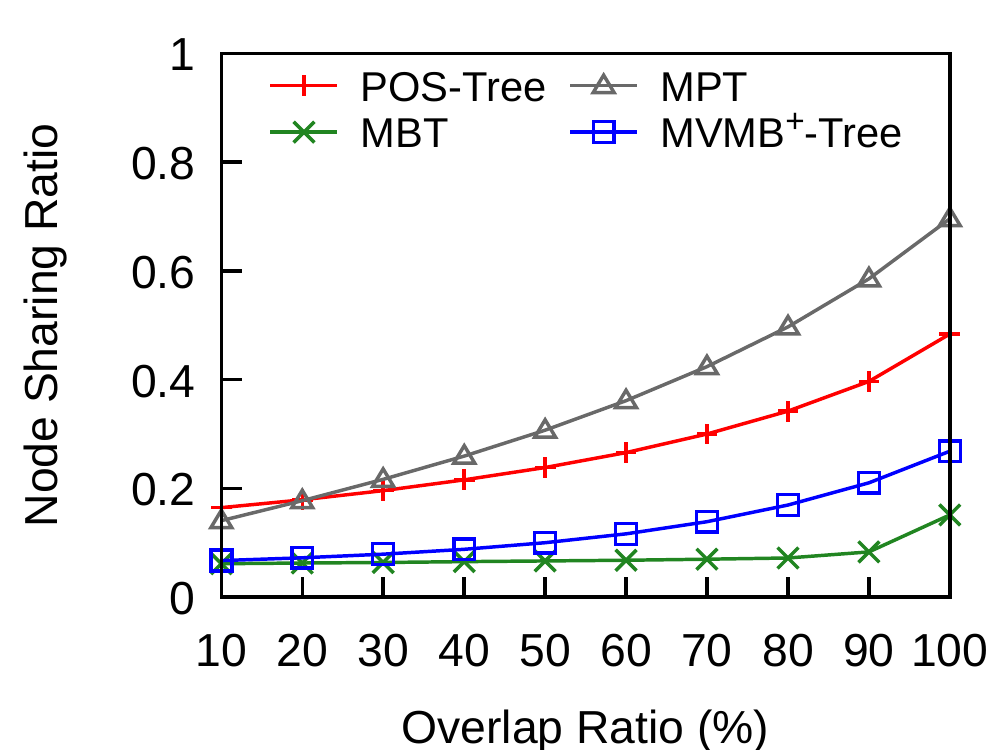}
    \label{fig:exp:overlap_share}
  }
  \caption{Performance on diverse group collaboration with varying overlap ratio}
\end{figure*}

\begin{figure*}
  \centering
  \subfigure[Storage]{
    \centering
    \includegraphics[scale=0.40]{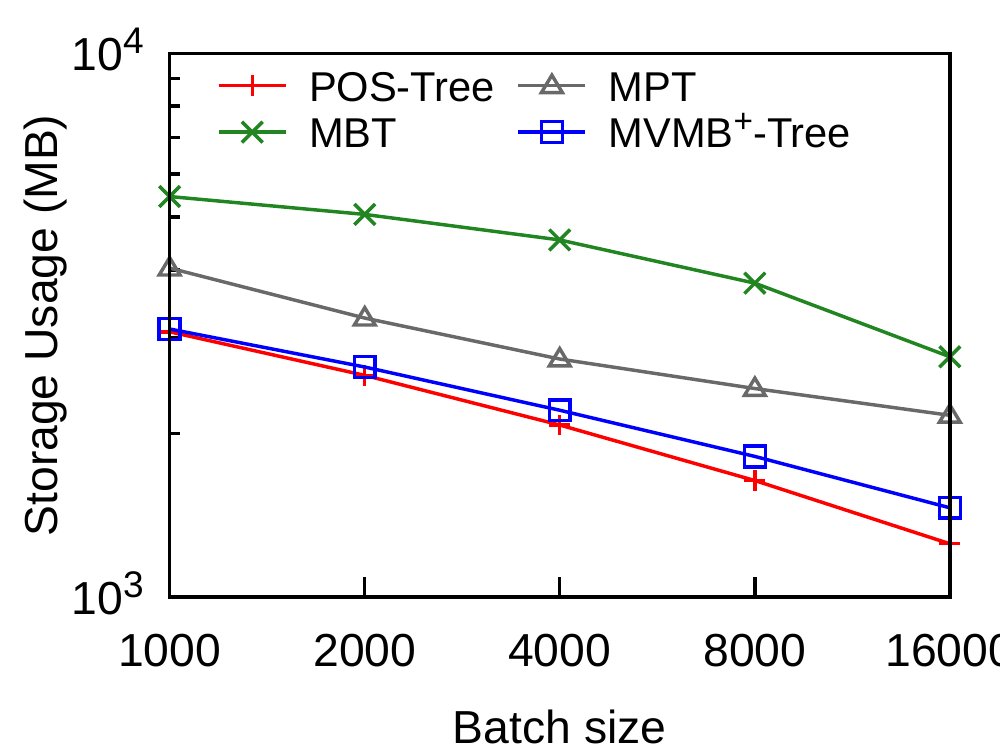}
    \label{fig:exp:batch_sto}
  }
  \subfigure[Number of nodes]{
    \centering
    \includegraphics[scale=0.40]{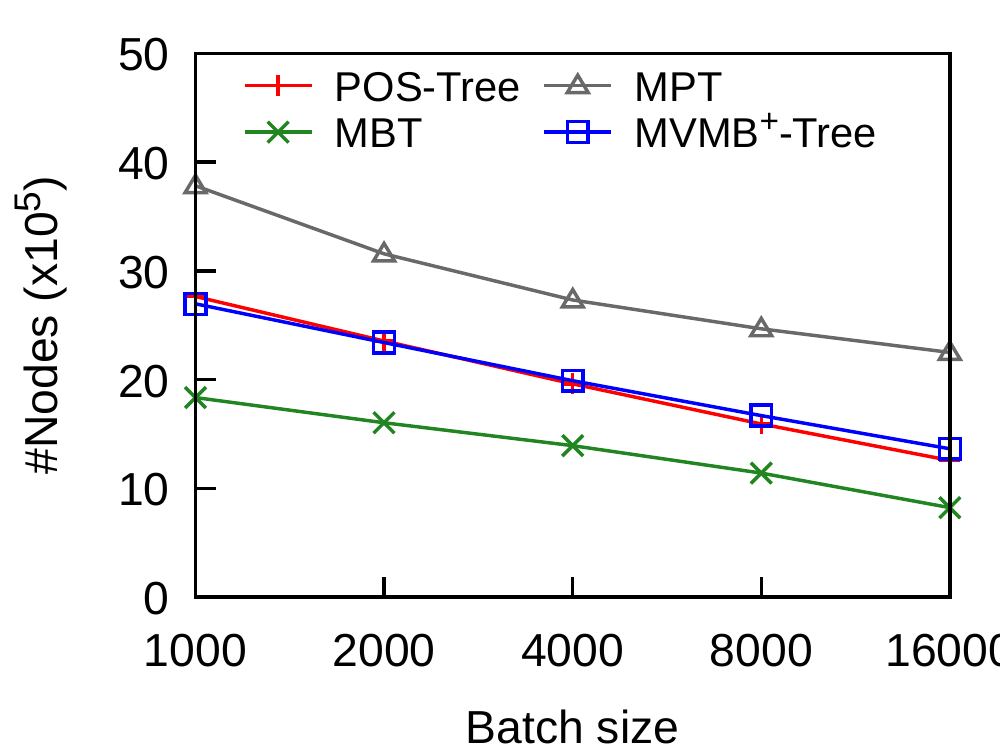}
    \label{fig:exp:batch_chk}
  }
  \subfigure[Deduplication ratio]{
    \centering
    \includegraphics[scale=0.40]{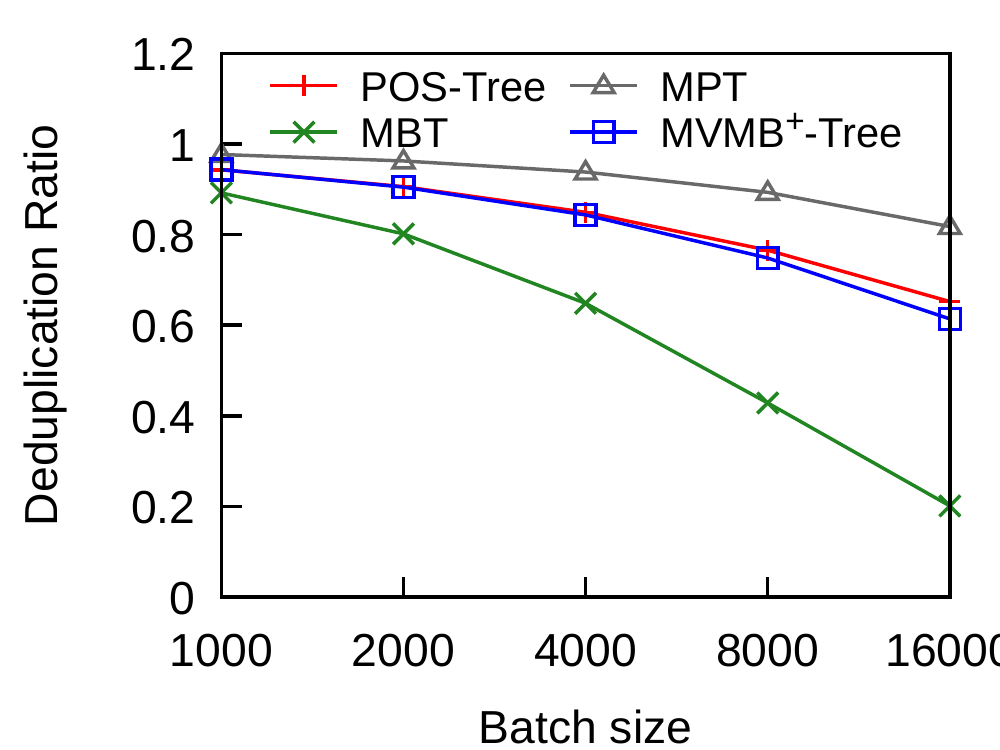}
    \label{fig:exp:batch_dedup}
  }
  \subfigure[Node sharing ratio]{
    \centering
    \includegraphics[scale=0.40]{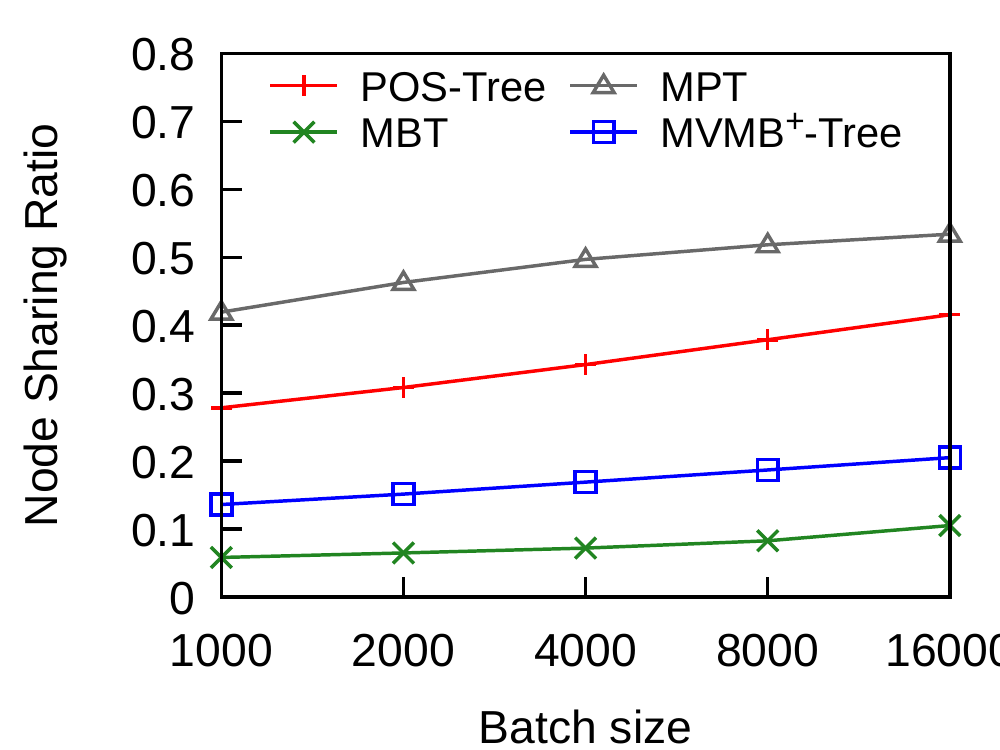}
    \label{fig:exp:batch_share}
  }
  \caption{Performance on diverse group collaboration with varying batch size}
\end{figure*}

To better analyze how the memory space is used by different pages, we further accumulate the number of nodes for all chosen indexes.
The results are demonstrated in Figure~\ref{fig:exp:basic_chk} with variant dataset sizes.
Typically, they follow similar trends as Figure~\ref{fig:exp:basic_sto}, except that MBT generates the least number of nodes as the total number of nodes is fixed for the structure.
The reason is rooted from the nature of MBT, which has a fixed total number of nodes and increasing leaf node size as more records are inserted.
Therefore, the number of nodes created keeps constant when updating or inserting, no matter how large the total number of records is.
On the contrary, other structures have a fixed node size and an increasing number of nodes, causing the number of nodes created, as well as the height of the tree, increases during updating or inserting.


The results for the Wiki dataset and Ethereum transaction dataset are shown in Figure~\ref{fig:exp:wiki_sto} and Figure~\ref{fig:exp:eth_sto}. 
Similar to the results of the YCSB experiment, MBT and MPT consumed more space than \mytree and MVMB$^+$-Tree. A difference is that MPT storage consumption increases very fast as the number of versions are loaded. This is because the key length of the Wiki dataset is much larger than that of YCSB, and the encoding method used by Ethereum further doubles the key length. This makes MPT a very sparse tree. For every insert/update operation, more nodes need to be re-hashed and created. Hence, the space efficiency is worse than it shows in the YCSB experiment.

Another difference is that MBT generates more nodes compared with other experiments. This is again because of the experiment setting that a new instance of index will be created per block. Since each block only contains a few hundreds of transactions, MBT is less efficient compared with other structures.

\subsubsection{Diverse Group Collaboration}
\label{subsubsec:storage-collaborative}
\noindent

We now examine the storage consumption in applications where different groups of users are collaborating to work on the same dataset.
This often occurs in the data cleansing process and data analysis procedure, where diverse parties work on different parts of the same dataset.
One significant phenomenon in this use case is that duplicates can be frequently found.
Therefore, the deduplication capability introduced by \myindex is critical to improving the space efficiency.
To evaluate the deduplication capability, we define another metric called node sharing ratio from a different aspect. The metric can be formulated as follows:
$$ \eta (S) = 1 - \frac{|P_1 \cup P_2 \cup ... \cup P_k|}{|P_1| + |P_2| + ... + |P_k|},$$
where $P_i$ is the set of nodes of an instance i. 
While the deduplication ratio evaluates the size of the storage saved, the node sharing ratio indicates how many duplicate nodes have been eliminated.

The YCSB dataset is used in this experiment.
We simulate 10 groups of users, each of which initializes the same dataset of 40,000 records.
We generate workloads of 160,000 records with overlap ratios ranging from 10\% to 100\% and feed them to the candidates.  
Here, 10\% overlap ratio means 10\% of the records have the same key and value.
The execution is processed with default batch size, i.e., 4,000 records.

The results of the deduplication ratio and the node sharing ratio are shown in Figure~\ref{fig:exp:overlap_dedup} and Figure~\ref{fig:exp:overlap_share}, respectively.
Both metrics of all the structures become higher when the workload overlap ratio increases since more duplicate nodes can be found due to increasing similarities among the datasets.
Benefiting from smaller node size and smaller portion of updating nodes, MPT achieves the highest deduplication ratio (up to 0.96) and node sharing ratio (up to 0.7).
\mytree achieves a slightly better deduplication ratio than the baseline though they both have similar size of nodes and the height of the tree.
The actual ratios are 0.88 and 0.86, respectively.
However, it achieves a much better node sharing ratio compared to the baseline (0.48 vs. 0.27) because of its content-addressable strategy when chunking the data.
By contrast, MBT's fixed number of pages and growing leaf nodes limit the number of duplicates, and therefore it does not perform as good as the other two \myindex representatives.

To be more precise, we further collect the storage usage and the number of pages created by the testing candidate and illustrate the results in Figure~\ref{fig:exp:overlap_sto} and Figure~\ref{fig:exp:overlap_chk}. 
The trends in the figures match the corresponding deduplication ratio and node sharing ratio perfectly.
With the increasing overlap ratio, storage reduction of \mytree and MPT is more obvious than the baseline, among which \mytree is the most space-efficient.
MPT is most sensitive to overlap ratio changes due to the high node sharing ratio introduced by its structural design.
Although it consumes more space for non-overlapping datasets, MPT outperforms the baseline for datasets with the overlap ratio above 90\%.

We also evaluate the effect of query batch size on storage space.
Same as previous experiments' setting, we simulate 10 parties.
Each of them initializes the same dataset contains 40,000 records and executes workloads containing 160,000 keys with default overlap ratio 50\%. 
Figure~\ref{fig:exp:batch_dedup} depicts how the deduplication ratio decreases along with the query batch size increases.
The reason is that larger batch sizes cause a larger portion of the index structure to be updated, resulting in fewer nodes to be reused between versions.
Figure~\ref{fig:exp:batch_sto} and Figure~\ref{fig:exp:batch_chk} show the storage usage and Number of nodes created with different batch sizes.
Similar relationships across the structures can be observed as in Figure~\ref{fig:exp:overlap_sto} and Figure~\ref{fig:exp:overlap_chk}. Except for both of the metrics decrease when using a larger batch size due to less versions stored in the index.

\subsubsection{Structure Parameters}
\noindent

The parameters of the indexes can affect the deduplication ratio as aforementioned in section~\ref{sec:analysis}.
The impact of those key parameters is verified in this experiment, namely node size for \mytree, number of buckets for MBT and mean key length for MPT.
For \mytree, the boundary pattern is varied to change the node size from 512 to 4,096 bytes probabilistically.
For MBT, the number of fixed buckets is set from 4000 to 10,000.
For MPT, the dataset is generated with different minimum key lengths, which can lead to diverse mean key length from 10.2 to 13.7.
(The maximum key length in the dataset is fixed.)
The results are shown in Table~\ref{table:pos}, which coincide with the conclusions in Section~\ref{sec:analysis}.
The deduplication ratio of \mytree increases as the average node size increases.
This is expected as the number of same large nodes is less than that of small nodes, leading to fewer occurrences of duplicate pages.
Similarly, the deduplication ratio of MBT increases as the number of buckets increases because a larger number of buckets results in smaller leaf nodes.
The deduplication ratio of MPT increases as the mean key length increases.
This is because longer keys usually have more conflicting bits and result in a wider tree.
Therefore, the portion of the reusable nodes increases.

\begin{figure*}
\begin{minipage}[b]{\textwidth}
    \centering
    \captionof{table}{Effect of structure parameters on the deduplication ratio}
    \begin{tabular}{|l|l|}
        \hline
        Node Size & $\eta$(POS-Tree) \\
        \hline
        512 & 0.722 \\
        \hline
        1024 & 0.6485 \\
        \hline
        2048 & 0.5391 \\
        \hline
        4096 & 0.4108 \\
        \hline
    \end{tabular}
    \hspace{2em}
    \begin{tabular}{|l|l|}
        \hline
        \# Buckets & $\eta$(MBT) \\
        \hline
        4000 & 0.3301 \\
        \hline
        6000 & 0.4599 \\
        \hline
        8000 & 0.5433 \\
        \hline
        10000 & 0.6003 \\
        \hline
    \end{tabular}
    \hspace{2em}
    \begin{tabular}{|l|l|}
        \hline
        $\overline{keylen}$ & $\eta$(MPT) \\
        \hline
        10.2 & 0.9685 \\
        \hline
        12 & 0.9693 \\
        \hline
        13.3 & 0.9806 \\
        \hline
        13.7 & 0.9823 \\
        \hline
    \end{tabular}
    \label{table:pos}
\end{minipage}
\end{figure*}

\begin{figure*}
  \centering
  \begin{minipage}{.49\textwidth}
    \subfigure[Deduplication ratio]{    
      \centering
      \includegraphics[scale=0.40]{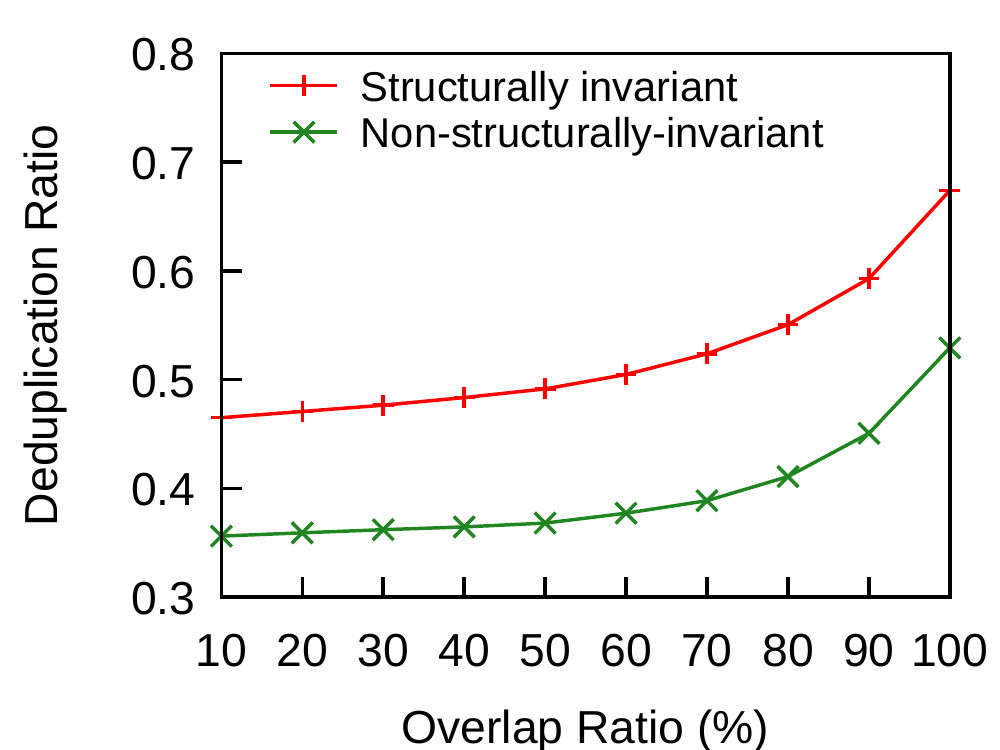}
      \label{fig:exp:nonsi_dedup}
    }
    \subfigure[Node sharing ratio]{
      \centering
      \includegraphics[scale=0.40]{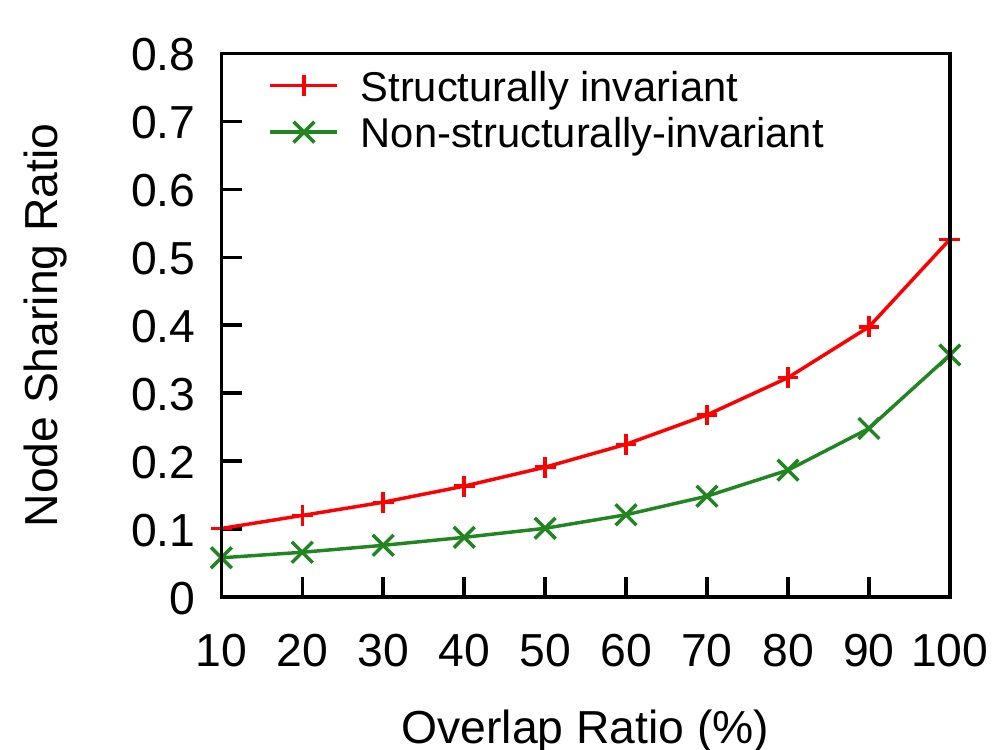}
      \label{fig:exp:nonsi_share}
    }
    \caption{Effect of Structurally Invariant property}
  \end{minipage}
  \begin{minipage}{.49\textwidth}
    \subfigure[Deduplication ratio]{    
      \centering
      \includegraphics[scale=0.40]{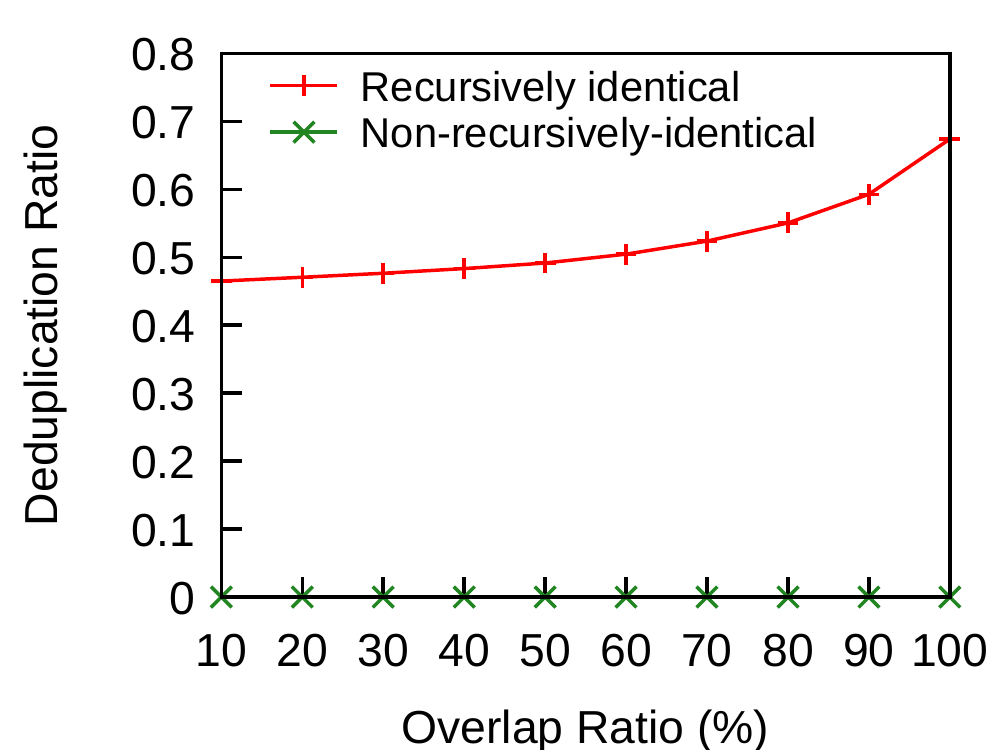}
      \label{fig:exp:nonri_dedup}
    }
    \subfigure[Node sharing ratio]{
      \centering
      \includegraphics[scale=0.40]{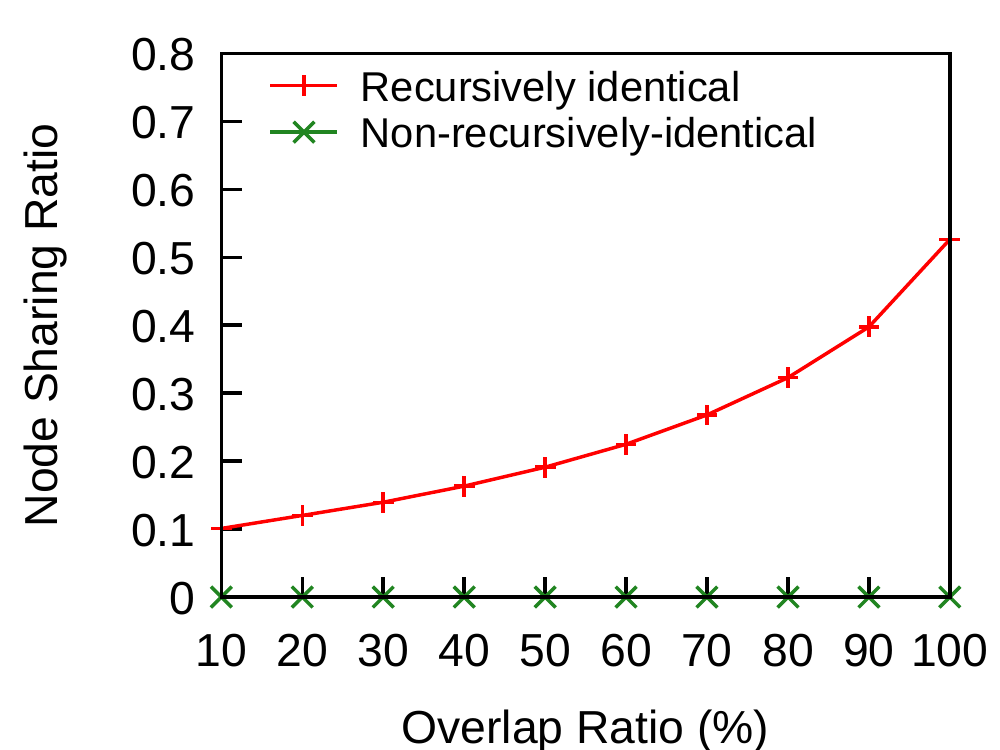}
      \label{fig:exp:nonri_share}
    }
    \caption{Effect of Recursively Identical property}
  \end{minipage}
\end{figure*}

\begin{figure*}[t]
  \begin{minipage}[t]{.49\textwidth}
  \centering
    \subfigure[Read]{
      \centering
      \includegraphics[scale=0.40]{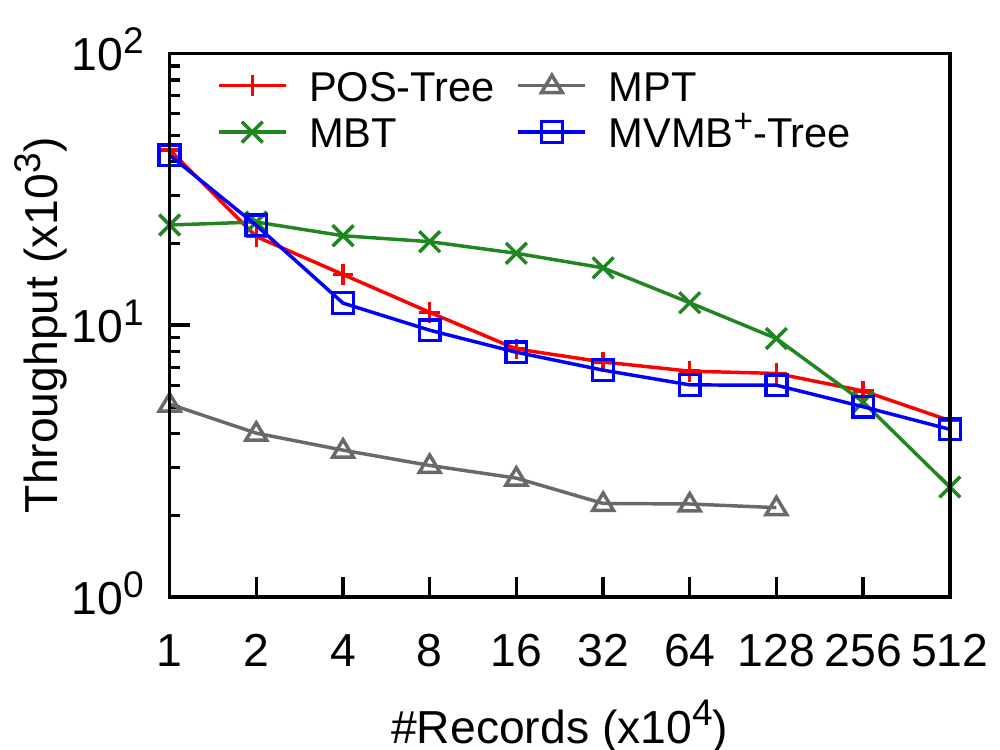}
      \label{fig:exp:sys_read}
    }
    \subfigure[Write]{
      \centering
      \includegraphics[scale=0.40]{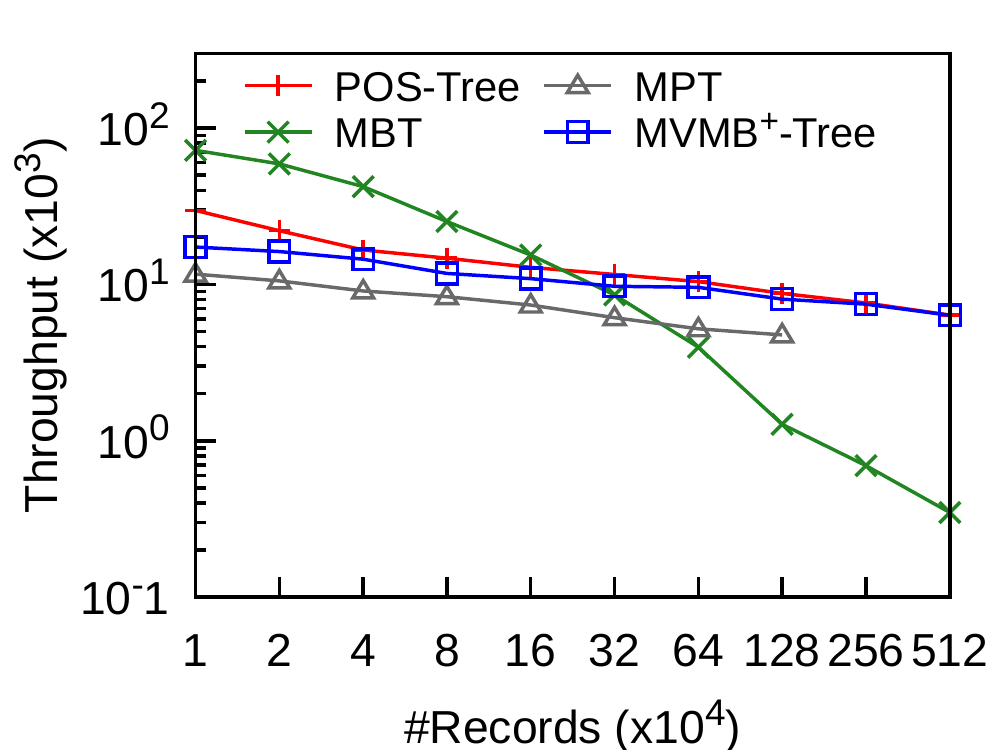}
      \label{fig:exp:sys_write}
    }
    \caption{Performance integrated with Forkbase}
    \label{fig:exp:sys}
  \end{minipage}
  \begin{minipage}[t]{.49\textwidth}
      \centering
      \subfigure[Read]{
        \centering
        \includegraphics[scale=0.40]{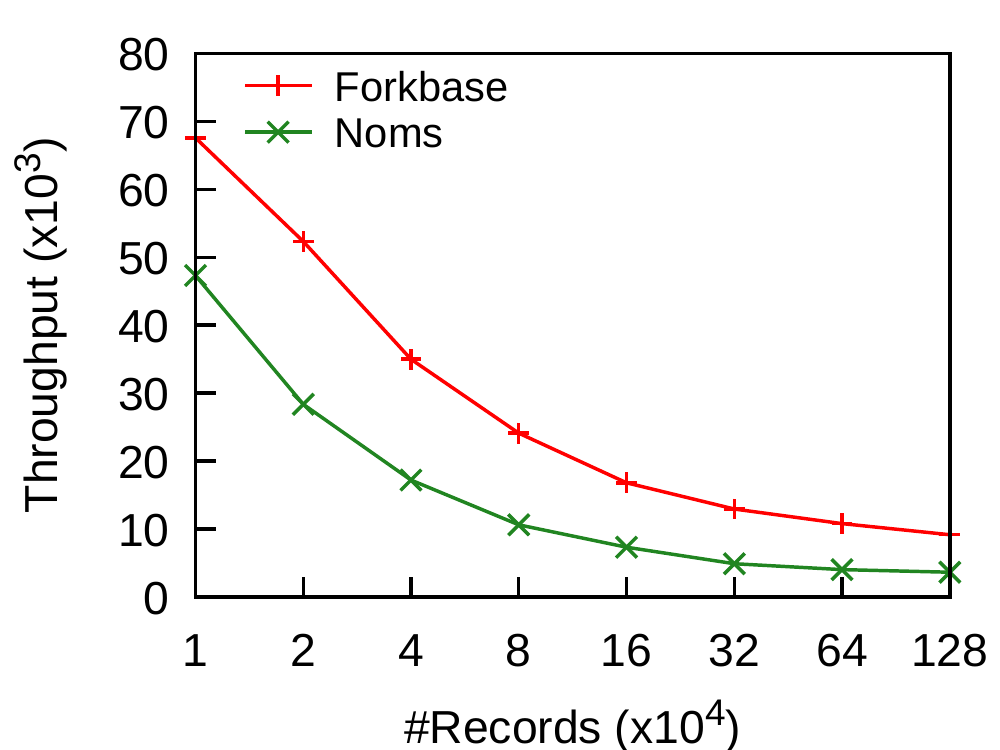}
        \label{fig:exp:fork_noms_read}
      }
      \subfigure[Write]{
        \centering
        \includegraphics[scale=0.40]{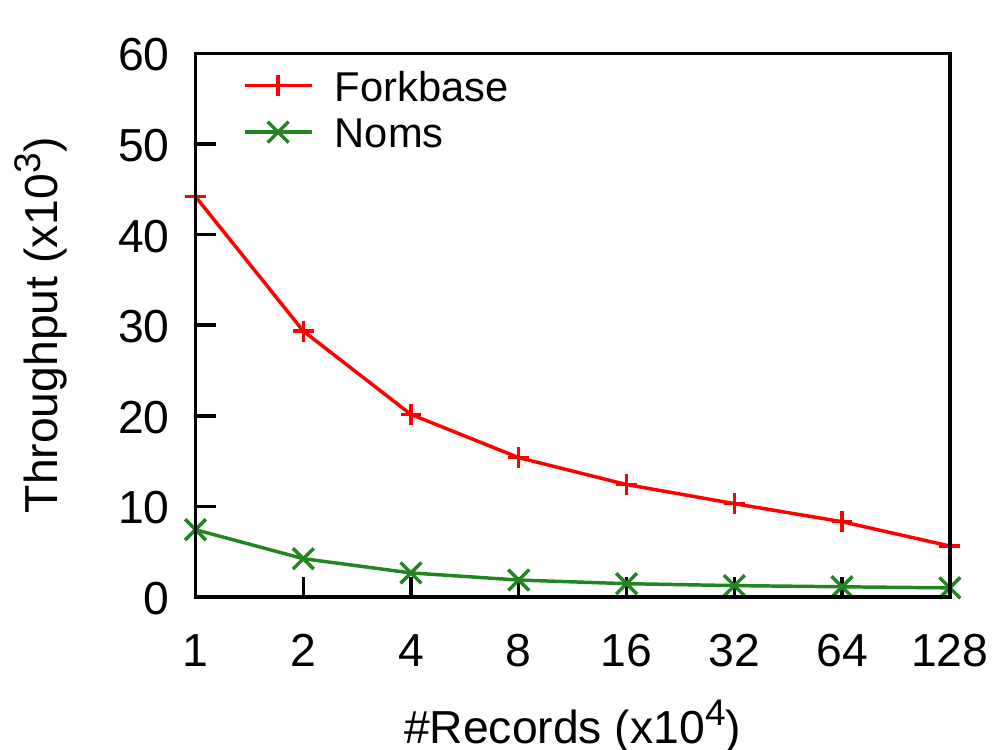}
        \label{fig:exp:fork_noms_write}
      }
    \caption{Comparison between Forkbase and Noms}
    \label{fig:exp:noms}
  \end{minipage}
\end{figure*}

\subsection{Breakdown Analysis}
\noindent

In this section, we evaluate how each \myindex property affects the storage and deduplication performance.
We select \mytree as the testing object and disable the properties one by one.
For each property, we first explain how each property is disabled and then provide the experimental results following closely.
We note that the \textit{Universally Reusable} property is common for all immutable tree indexes using copy-on-write approach. Thus, it is ignored in this experiment.

\subsubsection{Disabling Structurally Invariant Property}
\noindent


The pattern-aware partitioning is the key to guarantee the \textit{Structurally Invariant} property.
Therefore, we disable the property by forcibly splitting the entries at half of the maximum size when no pattern is found within the maximum size.
Consequently, the resulting structure depends on the data insertion order.
We increase the probability of not finding the pattern by increasing the bits of pattern and lowering the maximum value. 

The result is presented in Figure~\ref{fig:exp:nonsi_dedup}.
We can observe an up to 15\% decrease in the deduplication ratio when \textit{Structurally Invariant} property is disabled.
For instance, the deduplication ratio drops from 0.67 to 0.52 when the workload overlap ratio equals to 100\%.
It is expected as the index performs the operations in different orders, resulting in different nodes and smaller number of share-able pages.
Though the records stored are the same, \mytree cannot reuse the nodes with the \textit{Structurally Invariant} property disabled.
Similarly, Figure~\ref{fig:exp:nonsi_share} shows that the node sharing ratio decreases by up to 17\%, i.e. from 0.53 to 0.36, by disabling \textit{Structurally Invariant} property.


\subsubsection{Disabling Recursively Identical Property}
\noindent

Originally, only the set of nodes lying in the path from the root to the leaf node is copied and modified when an update operation is performed, while
the rest of the nodes are shared between the two versions in \mytree.
We disable \textit{Recursively Identical} property by forcibly copying all nodes in the tree.
The number of different pages between the two instances is much larger than the number of intersections, which is zero.

Figure~\ref{fig:exp:nonri_dedup} shows that the deduplication ratio for \mytree with \textit{Recursively Identical} disabled is 0 since the structure does not allow the sharing of nodes among different versions.
Obviously, the node sharing ratio of non \textit{Recursively Identical} \mytree shown in Figure~\ref{fig:exp:nonri_share} is also 0.
Compared to the figures in previous sections, we can infer how this property accelerates the deduplication rate and ultimately influences the final storage performance.

\vspace{1ex}

Overall, we can conclude that \textit{Recursively Identical} property is the fundamental property to enable indexes with deduplication and node sharing across different users and datasets.
On top of this, \textit{Structurally Invariant} property further enhances the level of deduplication and node sharing by making structures history-independent.

\subsection{System Experiment}
\noindent

\subsubsection{Integration with Forkbase}
\noindent

To further evaluate the performance of \myindex, we integrate the indexes into Forkbase~\cite{wang:2018}, a storage engine for blockchain and forkable applications.
In this experiment, we configured a single Forkbase servlet and a single client to benchmark the system-level throughput.
The evaluation results are demonstrated in Figure~\ref{fig:exp:sys}.

For read operations, the main difference between index-level performance and system integrated performance is the remote access due to client-server architecture. The overhead of remote access becomes the dominant factor of performance. To mitigate such overhead, Forkbase caches the nodes at clients after retrieved from servers. Hence, the following read operations on the same nodes can benefit from performing only local access.

Figure~\ref{fig:exp:sys_read} shows the throughput of read workload. Similar to index-level experiments, the throughput decreases when the total number of records grows.
POS-Tree achieves comparable performance to our baseline MVMB+-Tree, and it outperforms the other 2 indexes when the total number of records is large (greater than 2,560,000).
MPT performs the worst among all indexes due to larger tree height, which comply with the operation bound in Section~\ref{subsubsec:lookup}.
Different from the index-level experiment, MBT performs worse than POS-Tree and MVMB+-Tree when the number of records is extremely small (10,000 records).
This is because the hit ratio of cached nodes for MBT is lower than other indexes. 
Since all index nodes of MBT have a fixed number of entries, the number of repeated reads is less compared with POS-Tree and MVMB+-Tree, where large nodes contribute more repeated reads.
When the number of records grows larger, the affected portion of POS-Tree and MVMB+-Tree decreases.
Consequently, the number of repeated reads decreases.
While the structure of MVMB+-Tree keeps unchanged, leading to a constant number of repeated reads.
Therefore, MBT performs better when the number of records is greater than 20,000. 
When the number of records is greater than 2,560,000, the bottleneck becomes the loading time and the scanning time of leaf nodes, and the throughput drops below that of other indexes.

The write operations will be performed on the server side completely.
Hence they will not be affected by the hit ratio of cached nodes described above. 
Figure~\ref{fig:exp:sys_write} shows the throughput of write workload. We can observe similar results as that of index-level experiments. 

\subsubsection{Comparison between Forkbase and Noms}
\noindent

Next, we perform a comparative study between Forkbase and Noms~\cite{web:noms}.
Both systems facilitate data versioning management using similar indexing concept.
As in \mytree, the bottom most layer of Noms' Prolly Tree uses the sliding-window approach to partition the leaf nodes based on the boundary pattern.
To match the boundary pattern in the internal layers, \mytree directly uses hash values of the child nodes, while Prolly Tree uses the hash values repeatedly computed from the sliding-window.
Such computational overhead causes inefficiency of its write operations.

In the experiment, we directly use the code of Noms from its Github repository, which is implemented in GO.
We use Noms' remote setup on top of their own HTTP protocol to compare with Forkbase's single server single client setup as described previously.
To make a fair comparison, we configure the node size of POS-Tree to 4K with window size of 67 bytes, which is the default setting of Noms.
The experiment is conducted as follows. First, we initialize the systems with 10K to 128K records.
Then we execute read and write workload of 10K records respectively to measure the throughput.
The results are shown in Figure~\ref{fig:exp:noms}.
We can observe that Forkbase performs 1.4x-2.7x better in read operations and 5.6x-8.4x better in write operations than Noms.


\section{Conclusion}
\label{sec:conclu}
\noindent
Tamper evidence and deduplication are two properties increasingly demanded in emerging applications on immutable data, such as digital banking, blockchain and collaborative analytics.
Recent works~\cite{wang:2018,Wood:2014,web:hyperledger} have proposed three index structures equipped with these two properties. 
However, there have been no systematic comparisons among them.
To address the problem,
we conduct a comprehensive analysis of all three indexes in terms of both theoretical bounds and empirical performance.
Our analysis provides insights regarding the pros and cons of each index, based on which we conclude that \mytree~\cite{wang:2018} is a favorable choice for indexing immutable data.

\begin{acks}
	This research is supported by Singapore Ministry of Education Academic Research Fund Tier 3 under MOE's official grant number MOE2017-T3-1-007.
\end{acks}


\balance

\bibliographystyle{abbrv}
\bibliography{main-bibliography}

\end{document}